\documentclass[a4paper,fleqn,usenatbib]{mnras}

\usepackage{newtxtext,newtxmath}
\usepackage{enumitem}

\usepackage[T1]{fontenc}
\usepackage{ae,aecompl}
\usepackage{multirow}

\usepackage{graphicx}	
\usepackage{amsmath}	

\usepackage{color}
\usepackage{BibDef}
\def\lsim{\mathrel{\rlap{\lower 3pt \hbox{$\sim$}} \raise 2.0pt \hbox{$<$}}}
\def\gsim{\mathrel{\rlap{\lower 3pt \hbox{$\sim$}} \raise 2.0pt \hbox{$>$}}}
\def\msun{\rm\, {M_\odot}}
\def\kms{\rm\, km\,s^{-1}}

\def\gizmo{\textsc{gizmo}}

\title[High-redshift QSOs] 
{High-redshift quasars and their host galaxies II: multiphase gas and stellar kinematics}
\author[A. Lupi et al.]{Alessandro Lupi,$^{1,2}$\thanks{E-mail:
alessandro.lupi@unimib.it} Marta Volonteri,$^3$ Roberto Decarli,$^4$ Stefano Bovino,$^5$\newauthor and Joseph Silk$^{3,6,7}$\\
$^1$Dipartimento di Fisica ``G. Occhialini'', Universit\`a degli Studi di Milano-Bicocca, Piazza della Scienza 3, I-20126 Milano, Italy\\
$^2$INFN – Sezione di Milano-Bicocca, Piazza della Scienza 3, I-20126 Milano, Italy\\
$^3$Sorbonne Universit\`{e}s, UPMC Univ Paris 6 et CNRS, UMR 7095, Institut d'Astrophysique de Paris, 98 bis bd Arago, F-75014 Paris, France\\
$^4$INAF -- Osservatorio di Astrofisica e Scienza dello Spazio di Bologna, via Gobetti 93/3, I-40129, Bologna, Italy\\
$^5$Departamento de Astronomía, Faculdad Ciencias Físicas y Matemáticas, Universidad de Concepción, Av. Esteban Iturra s/n Barrio Universitario, \\Casilla 160, Concepción, Chile\\
$^6$Department of Physics and Astronomy, The Johns Hopkins University, Baltimore, MD 21218, USA\\\
$^7$BIPAC, University of Oxford,1 Keble Road, Oxford OX1 3RH, UK
}

\begin{document}

\date{Draft \today}

\pagerange{\pageref{firstpage}--\pageref{lastpage}} \pubyear{2019}

\maketitle

\label{firstpage}

\begin{abstract}
Observations of $z \gtrsim 6$ quasars provide information on the early phases of the most massive black holes (MBHs) and galaxies. Current observations at sub-mm wavelengths trace cold and warm gas, and future observations will extend information to other gas phases and the stellar properties. The goal of this study is to examine the gas life cycle in a $z \gtrsim 6$ quasar: from accretion from the halo to the galaxy and all the way into the MBH, to how star formation and the MBH itself affect the gas properties.
Using a very-high resolution cosmological zoom-in simulation of a $z=7$ quasar including state-of-the-art non-equilibrium chemistry, MBH formation, growth and feedback, we investigate the distribution of the different gas phases in the interstellar medium across cosmic time. We assess the morphological evolution of the quasar host using different tracers (star- or gas-based) and the thermodynamic distribution of the MBH accretion-driven outflows, finding that obscuration in the disc is mainly due to molecular gas, with the atomic component contributing at larger scales and/or above/below the disc plane. Moreover, our results also show that molecular outflows, if present, are more likely the result of gas being lifted near the MBH than production within the wind because of thermal instabilities. Finally, we also discuss how different gas phases can be employed to dynamically constrain the MBH mass, and argue that resolutions below $\sim 100$~pc yield unreliable estimates because of the strong contribution of the nuclear stellar component to the potential at larger scales.
\end{abstract}
\begin{keywords}
quasars: supermassive black holes - galaxies: ISM - galaxies: formation - galaxies: evolution.
\end{keywords}

\section{Introduction}

Massive black holes (MBHs) are ubiquitously detected in the centre of massive galaxies at all redshifts, with masses going from about $10^5\msun$ up to $10^{9-10}\msun$. Observationally, they can be identified via gas accretion, when a fraction of accreting material energy is released in the form of radiation, turning them into active galactic nuclei (AGN), or when powerful collimated jets are launched. According to  Soltan's argument \citep{soltan82}, most of the mass of these MBHs have been gained via accretion, hence they should have formed as smaller `seeds' when the Universe was much younger.

A strong constraint on the seed formation model and the MBH early growth comes from the observations of high-redshift galaxies at $z\gtrsim 6$ (when the Universe was less than 1~Gyr old) hosting MBHs with masses around 10$^9\msun$ \citep{fan06,mortlock11,banados18}. It is therefore crucial to properly resolve the environment around these objects and the properties of their galaxy hosts in order to improve our understanding of how these MBHs formed and managed to grow so quickly.

Currently, observations sampling the rest-frame far infrared (FIR; in particular the [CII] line at 158~$\mu$m and the CO line emission), that are uncontaminated by the quasar light and unaffected by dust extinction, have shaped our understanding of quasar host galaxies at cosmic dawn. Thanks to maps of these tracers, now resolved to kpc scales in tens of $z\gtrsim 6$ quasar host galaxies with ALMA (see, e.g., \citealt{venemans19, venemans20, neeleman21}), we are getting important information about the kinematics and dynamics of molecular gas within the quasar hosts. Despite the rapid progress over the last few years, however, the multiphase interstellar medium (ISM) structure in high-redshift systems is still poorly constrained. Critical open questions concern the interplay between gas inflows and outflows; the energy balance of the cold medium and its connection with star formation; and the gas kinematics in the immediate vicinity of the black holes.

Fast nuclear (sub-pc) and galaxy-scale (kpc) outflows have been observed mostly at low/intermediate redshifts, allowing us to infer the AGN impact on its host \citep{feruglio10,greene12,tombesi13,cicone14,harrison14,nardini15}. At high redshift, such strong outflows have only been observed in a few cases \citep[e.g.][]{maiolino12,cicone15,stanley2019}. Ultrafast outflows approaching 10 per cent of the speed of light have been also detected in the X-ray band \citep{gofford15}, in 40 per cent of the bright local AGN population. Recently, cold molecular winds and outflows (up to $10^{10}\msun$ in molecular gas) have been also found, with velocities around $\sim$10$^2$--10$^3$~km~s$^{-1}$ \citep[e.g.][]{fischer10,cicone14,garcia-burillo14,combes17,cicone18,sirressi19}. Understanding the impact of outflows in regulating the build-up of gaseous reservoirs, the formation of stars and the nuclear activity is of paramount importance in order to assess how these first massive galaxies evolve.

Another important open issue in the study of high-redshift galaxies is the reliability of FIR emission as a tracer of the cold gas mass and the star formation process. The FIR dust continuum emission is sustained by the reprocessing of UV light predominantly arising from young stars. Thus, it is often used as a tracer of the star formation rate (SFR). In local spiral galaxies, the [CII] line luminosity appears to correlate with the FIR luminosity (see, e.g., \citealt{malhotra01}), which justifies its use as a SFR tracer \citep{delooze14, herreracamus18}. However, [CII] emission may appear suppressed, either because of the so-called [CII] deficit, which leads to underluminous [CII] at a given FIR luminosity in the presence of compact starbursts \citep[e.g.,][]{diaz-santos17}, or because of a low metallicity ISM \citep{vallini15,capak15,carniani18,lupi20} and different ionisation field compared to low-redshift galaxies \citep[see, e.g.][]{arata20}. A related question concerns the large amount of dust ($M_{\rm dust}\sim 10^8\msun$) observed in all quasars above $z=6$, that requires efficient dust formation mechanisms. While several studies tackling this problem have already been performed \citep[see, e.g.][]{valiante11,graziani20}, a proper answer is still to be found, in part due to the uncertainties in the modelling of dust evolution in theoretical models, but also resulting from the crude approximations made in the modelling of the ISM \citep[a limitation some recent studies tried to overcome; see, e.g.][]{pallottini17b,pallottini19,lupi19b}.

Because the quasars studied so far at cosmic dawn were selected based on enormous energy release of the accreting black holes, accurate measures of the MBH mass is pivotal. 
ALMA is in principle able to spatially resolve the cold gas dynamics within the sphere of influence of the MBH in these quasars \citep{venemans19}, thus providing very robust MBH mass estimates as done in local galaxies \citep[see, e.g.,][]{davis18}. This allows for an independent check on the MBH estimates based on optical/UV emission line widths \citep{vestergaard08} and to investigate more thoroughly the relation between MBH and galaxy properties, with the caveat of  uncertainties in the dynamical mass estimates \citep[Paper~I,][]{lupi19b} and biases related to the  selection based on quasar luminosity  \citep{lauer07}.

From a modelling perspective, many authors have recently addressed the evolution of high-redshift quasar hosts with different techniques, i.e. semi-analytic models \citep[e.g.][]{valiante11,valiante14,pezzulli16} and numerical simulations \citep{costa14,richardson16,dimatteo17,smidt18,barai18,lupi19b}, with the aim of addressing different questions. For instance, \citet{dimatteo17} and \citet{ni20} focus on the conditions for the efficient growth of these MBHs and on their obscuration, while \citet{costa14} and \citet{smidt18} assess the star formation (SF) suppression by AGN radiative and mechanical feedback. \citet{richardson16} and \citet{barai18} investigate the impact of AGN feedback in a $z=5$ proto-cluster of galaxies and in a $z=6$  halo of $10^{12}\msun$, respectively, while \cite{costa15} and \cite{ni2018} study outflows in $z>6$ quasars \citep[see also][for a fainter AGN]{prieto17}. However, the limited mass/spatial resolution in these works does not allow for a proper treatment and description of the thermodynamic evolution of the dense ISM, and more sophisticated and higher resolution simulations are needed. In this work, instead of looking at the large scales around many quasar hosts, we opt for focussing on a single galaxy at an unprecedented level of detail in terms of physical modelling: non-equilibrium chemistry coupled with extremely high spatial and mass resolution, so that phase changes of the gas are accounted for naturally down to molecular cloud scales.

This is the second of a series of paper addressing properties of high-redshift quasar hosts and their MBHs. In paper I, we presented and discussed the main evolution of the target galaxy and its central MBH, focussing on the stellar and gas tracers (total gas and [CII] emission). Here, we extend the analysis of the main galaxy including molecular hydrogen, directly traced in the simulation, and focussing on the dynamics and morphology of the main galaxy as a function of redshift. In Paper III (Lupi et al. in prep.) we will discuss the evolution of the entire MBH population forming during the simulation.

The paper is organised as follows: in Section~\ref{sec:setup}, we recap the setup of the simulation; in Section~\ref{sec:multiphase}, we present our results; in Section~\ref{sec:conclusions}, we draw our conclusions.

\section{Simulation setup}
\label{sec:setup}
The simulation presented in paper I followed the evolution of a massive halo ($M_{\rm halo} \sim 3\times 10^{12}\msun$ at $z=6$) expected to represent a quasar host \citep{dimatteo17}. The initial conditions were accurately created to match the expected halo mass \citep{dimatteo17,tenneti18} and the galaxy overdensity significance \citep{uchiyama18,mignoli20} via \textsc{music} \citep{hahn13}, adopting the \citet{planck16} cosmological parameters, with $\Omega_{\rm m}=0.3089$, $\Omega_\Lambda = 0.6911$, $\Omega_{\rm b}=0.0489$, $\sigma_8 = 0.8159$, $n_{\rm s} = 0.9667$, and $H_0 = 67.74\,\rm km\, s^{-1} Mpc^{-1}$, with no contribution from radiation and curvature.
In order to reach the desired resolution, a Lagrangian volume extending up to 2.5 virial radii of the target halo was recursively refined following the approach by \citet{fiacconi17} to ensure that no contamination by low-resolution dark matter particles was present within the virial radius.

The simulation has been run with \textsc{gizmo} \citep{hopkins15}, a particle-based code descendant of \textsc{Gadget3} and \textsc{Gadget2} \citep{springel05}, employing the meshless-finite-mass method, a fully Lagrangian scheme which exhibits excellent shock-capturing properties and almost perfect conservation of angular momentum. 
The spatial resolution of the simulation is 40 and 10~pc h$^{-1}$ for dark matter and stars, respectively, and fully adaptive softening down to a minimum of $\sim 5$~pc is assumed for gas particles/cells. The mass resolution is $\sim 10^4\msun$ for baryons and $\sim 10^5\msun$ for dark matter.

Our simulation was performed with state-of-the-art sub-grid prescriptions that allowed us to follow in detail non-equilibrium chemistry of primordial species, as in \citet{lupi18}, star formation and stellar feedback, as in \citet{lupi19a}, and MBH seeding, accretion, and feedback \citep[see][for details]{lupi19b}.
In particular, the gas thermodynamics is self-consistently followed via the non-equilibrium chemistry library \textsc{krome} \citep{grassi14}, directly coupled to the hydrodynamics and the radiation feedback model in the code, following time-dependently the chemical evolution (abundances) of nine primordial species (H, H$^+$, H$^-$, He, He$^+$, He$^{++}$, H$_2$, H$_2^+$, and e$^-$) under the effect of multiple heating and cooling processes \citep[e.g. atomic cooling, molecular cooling, chemical cooling and heating, photoheating, dust cooling, photoelectric heating, H$_2$ UV pumping;][]{lupi18,lupi19b}. This approach naturally follows all the phase changes in the gas (ionised/neutral/molecular), without applying any ad-hoc cut in temperature, and also accounts for intrinsic multi-phase distribution within gas cells.
Star formation is based on the multi-free fall model of \citet{padoan11}, as described in \citet{federrath12}, in which the average star formation efficiency per molecular cloud is a function of the cloud properties, in particular the Mach number and the virial parameter, which define how bound and turbulent the cloud is. In our simulation, we assume that each gas particle/cell corresponds to an entire or a significant portion of a molecular cloud, and use the star formation efficiency just described to rescale the probability of the particle to be converted into stars. When a star particle is formed, we assume it represents an entire stellar population with a \citet{chabrier03} initial mass function and a fixed metallicity (inherited from the gas progenitor). As the population gets older, massive stars explode as supernovae (SNe), which we model via an instantaneous energy injection of $10^{51}$~erg per SN. The SN energy is distributed among gas neighbours both in thermal and kinetic form, with the ratio between the two components determined according to the results by \citet{martizzi15}. Stellar radiative feedback is implemented in an approximate fashion, collecting the radiative flux reaching each gas particle in the same way gravity is calculated, i.e. via the gravity tree, assuming attenuation only around the stellar source and at the target, and an optically thin medium in-between.\footnote{This has been shown to well reproduce the results of on-the-fly radiative transfer calculations \citep{lupi18} at a much cheaper cost.} MBHs are seeded in galaxies with more than $10^8\,\msun$ in stars, and are allows to accrete from their surrounding according to the Bondi-Hoyle-Lyttleton formula \citep{bondi52}. The accretion rate is determined as in \citet{choi12}, while the accretion-powered luminosity is released to the neighbouring gas in a purely thermal fashion, assuming a radiative efficiency of 0.1 \citep[see][for details]{lupi19b}.
As in paper I, the halo is identified using \textsc{amiga halo finder} \citep{knollmann09}, and the analysis is based on the particles within a sphere of radius $0.2 r_{\rm vir}$ only, to exclude contribution from satellite galaxies.

\section{The multiphase gas in the quasar host}
\label{sec:multiphase}
Besides the stellar content,  the various phases of the ISM are also traced directly in our simulation. This allows us to, on the one hand, study physically how the multiphase structure evolves, and, on the other hand, to test how accurately tracers can reconstruct this evolution. We analyse the simulation down to $z=7$, first looking at the intrinsic properties, and then considering different observational tracers.

\begin{figure}
\includegraphics[width=\columnwidth]{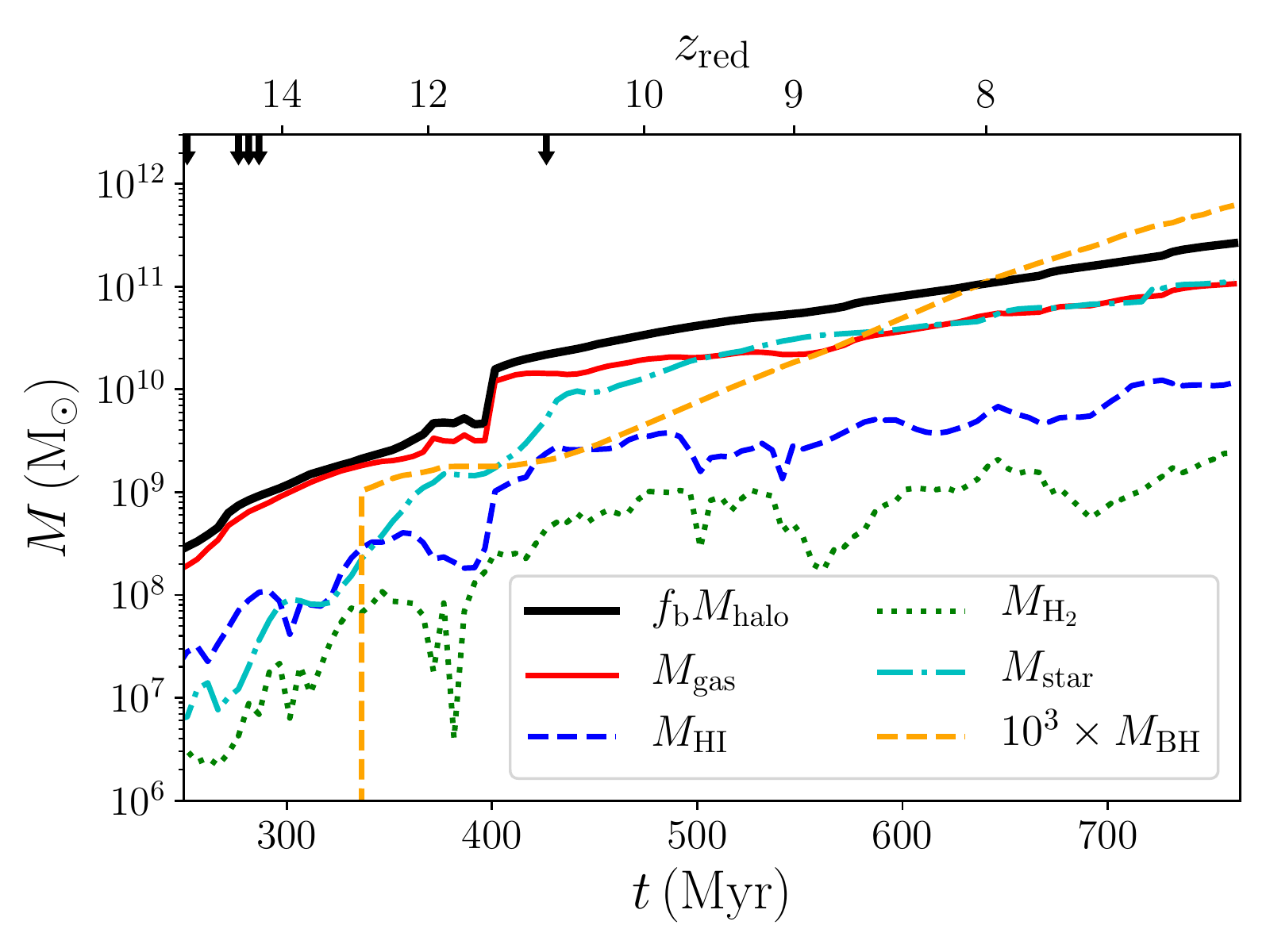}
\caption{Redshift evolution of the mass of different gas phases, compared with the stellar mass (cyan dot-dashed line) and the BH mass (orange dashed line), the latter scaled up by a factor of 1000. We show the universal baryonic mass as a black thick solid line, the total gas mass within the halo as a red solid line, and the neutral/molecular gas masses as blue dashed and green dotted lines. The black arrows correspond to the epochs during which the stellar mass grows by more than 50\% in 5~Myr. Compared to the total gas mass available in the halo, that strictly follows the halo growth in mass, only 10 per cent of it is able to cool down and form the HI disc of the galaxy. An even smaller fraction, about 1 per cent, becomes dense enough to form H$_2$, which then forms stars. Interestingly, the fractions of HI and H$_2$ remain almost constant relative to the total gas available, with only small fluctuations due to star formation, that remove cold and dense gas, and stellar feedback, that evacuates gas from dense gas and pushes it in the outer halo (or even out of it), turning into hot ionised gas.} 
\label{fig:mass_z}
\end{figure}

\begin{figure}
\includegraphics[width=\columnwidth,trim=1cm 1.3cm 1cm 1.5cm,clip]{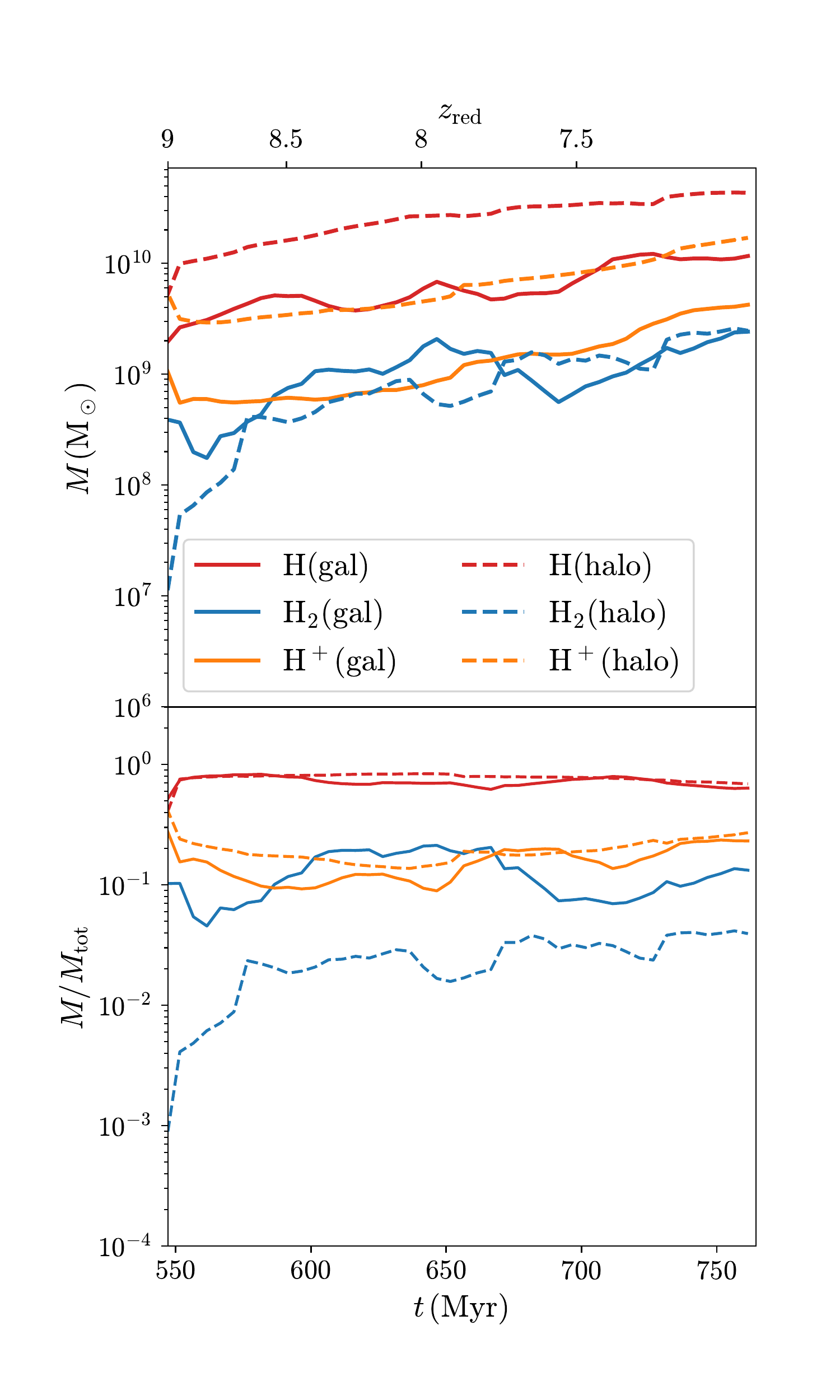}
\caption{Redshift evolution of the mass (top panel) and mass fraction (bottom panel) of different gas phases in the galaxy host (solid lines) and in the halo (dashed lines). Orange, red, and blue lines correspond to the ionised, atomic, and molecular phases respectively. Within the central region most of the mass is in the atomic phase, with about 10\% in the molecular/dense and ionised phases. In the halo, instead, the total mass in the atomic and ionised phases is typically larger, despite the lower densities, whereas the molecular phase is smaller than (or at most comparable to that) in the galaxy. In terms of fractions, atomic and ionised gas phases do not show significant differences between the galaxy and the halo, unlike the molecular phase, which is strongly subdominant in the halo. } 
\label{fig:phases_evol}
\end{figure}
\subsection{Intrinsic properties}

\subsubsection{Mass evolution}

In the local Universe star-forming massive galaxies generally have a low molecular gas fraction \citep[$\sim 10\%$,][]{saintonge17}, while in $z\sim 2$ massive galaxies the molecular gas content is similar to the stellar fraction, with an empirical scaling with redshift $(1+z)^{2.71}$ for galaxies on the main sequence of star formation \citep{genzel15,aravena19}. In Paper~I we have compared the total gas mass to the stellar mass, but we can now look at the various gas phases, including the molecular gas.

 In Fig.~\ref{fig:mass_z}, we show the evolution of the different gas phases with redshift, as blue dashed (HI) and green dotted (H$_2$) lines, directly estimated from the mass fractions of each species associated to each gas particle (see Section~\ref{sec:setup} for details about how the species are tracked). The total gas mass within the halo ($r<r_{\rm vir}$) is shown as a red solid line, and the universal baryon fraction as a black thick solid line. For comparison, we also show as a cyan dot-dashed line, the stellar mass, and as a dashed orange line, the BH mass, scaled up by a factor of 1000. Because of star formation (SF), a large fraction of gas is converted into stars, and the available gas mass settles on a 1:1 ratio with the stellar mass, as shown in Paper I. Relative to the total available gas mass, only 10 per cent is able to cool and recombine into HI, settling in a proto-galactic disc, and an even smaller fraction ($f_{\rm H_2}\sim 0.01$) becomes dense and cold enough to form H$_2$, providing fresh material for SF. As time elapses, the total amount of gas in the different phases evolves coherently, with almost constant ratios of about 0.1 and 0.01 for HI and H$_2$, respectively. This is consistent with the results by \citet{walter20}, where the gas reservoir mass evolution in the range $z\sim 4-0$ seems to be directly set by the gas inflow onto the halo, minus the contribution of star formation. Nevertheless, large fluctuations can be observed, especially for H$_2$, corresponding to  bursts of SF, which consume dense and cold gas, and supernovae (SNe), that disperse and heat up gas, turning it back into the neutral or even ionised phase.  

The molecular-to-stellar mass fraction stays in the range 0.02-0.07, suggesting that our quasar host, despite being on the star formation main sequence (see Paper~I) is not following such a strong evolution in the relationship with redshift as suggested by lower-redshift observations of star-forming galaxies.

In order to disentangle how the different phases are distributed in the quasar host, we show in Fig.~\ref{fig:phases_evol} the masses (top panel) and the mass fractions (bottom panel) in the different phases in the galaxy host (identified as the region enclosed within a sphere of radius $R_{\rm centre}=\min\{0.2R_{\rm vir},10\rm\, kpc$\}) and in the halo (represented by the spherical shell extending from $R_{\rm centre}$ up to $R_{\rm vir}$). In the galaxy region, most of the gas is in the atomic phase, with the molecular/dense and ionised phases representing at most 10\%. 

These ratios are the result of the typically high gas densities in this region, in which efficient cooling allows for rapid gas recombination into the neutral phase, whereas the subsequent conversion into the molecular phase is less efficient because of a combination between weaker cooling and the dissociating/ionising effect of stellar feedback (via radiation or SNe sweeping away the gas). In the halo, the gas is still mainly atomic, with ionised gas only contributing to 1/10th of the total mass. On the other hand, the molecular component only represents a tiny fraction of the total mass (lower or at most comparable to the mass within the galaxy), and is associated with clumps along gas streams, able to cool down and to form H$_2$, and dense gas within galaxy satellites around the quasar host. With time, the mass in the different phases grows following the halo mass growth, in both regions. However, while the growth in the halo is  smooth (but for the molecular phase), the increase in neutral and molecular gas within the galaxy exhibits significant fluctuations, due to the interplay between gas cooling, star formation, and stellar feedback, as can be inferred from Fig.~\ref{fig:mass_z} by comparing the evolution of the H$_2$ mass with that of the stellar mass. 

To better understand how important different phases are in the evolution of the galaxy, we analysed the relative mass variation with time of the different phases in the galaxy and in the halo, by randomly selecting $10^4$ particles per snapshot, and tracking them forward in time to the next snapshot, estimating the mass in each phase and the mass converted into stars. We found that, in the halo, the phase evolution is slow, with very small variations ($<10\%$) but for the molecular phase, which is extremely sensitive to the different physical processes (star formation, stellar, and AGN feedback). In the galaxy, instead, the variations are larger, and reach up to 20-30\%, with mild fluctuations except for the molecular gas again, which exhibits strong fluctuations from positive (due to cooling and H$_2$ formation) to negative (due to dissociation/ionisation) values.

The atomic and ionised mass variation points to a systematic decrease, which can be explained by the fact that a lot of warm/hot gas recombines because of cooling, and part of it, after passing to the atomic phase, contracts and converts into molecular gas. Although this seems to suggest that the atomic phase is simply a transition stage, the timescale over which the transition to molecular gas occurs is quite long, hinting for a persistent atomic phase in the galaxy that lasts for relatively long times. Finally, about 20\% of the mass sampled gets converted into stars within 5~Myr, most of it corresponding to  molecular gas already present in the galaxy, but part also resulting from the conversion of atomic gas (red lines) into H$_2$ and quickly into stars, as can be noticed by comparing the red and green lines.

\subsubsection{Size evolution}

Large simulations have shown that statistically the stellar component of the host galaxies of luminous high-z quasars is expected to be very compact \citep{tenneti18b} and bulge-dominated \citep{marshall19}, barely resolvable by JWST. Given our high spatial resolution, $\sim $ 10 (100) times higher for star (gas) than in large-volume simulations, we can  resolve both the scale length and the scale height of the galactic disc and assess more accurately its extent, although on a single galaxy. Thanks to our chemical network we can further investigate and compare the evolution of the physical size of several galaxy components: stars, neutral hydrogen, molecular hydrogen, star-forming gas. In particular, different instruments have different resolution and we can compare high and low resolution images of the same galaxies.

\begin{figure}
\includegraphics[width=\columnwidth]{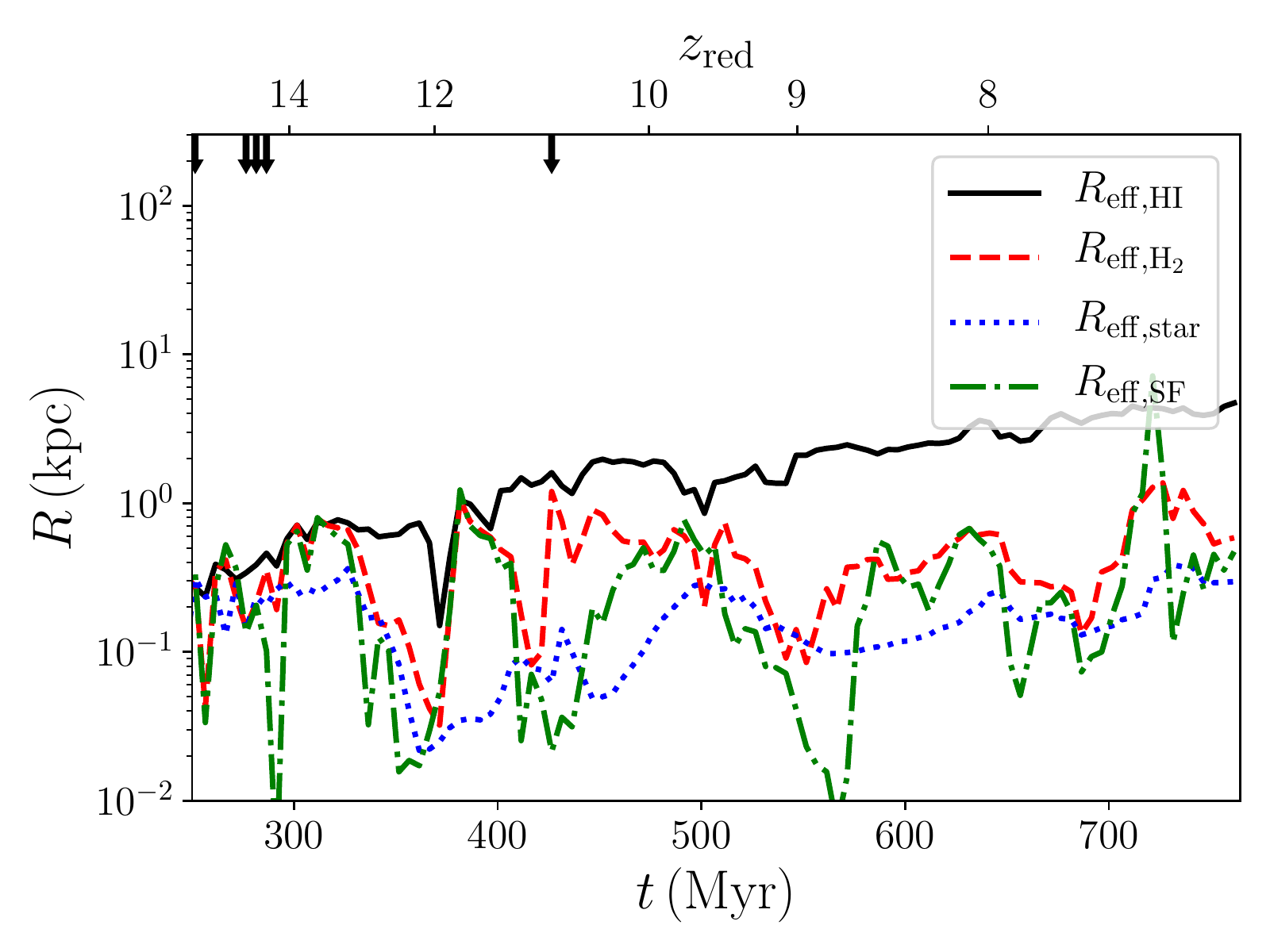}
\caption{Redshift evolution of the half-mass radius for the neutral hydrogen component (black solid line), H$_2$ (red dashed line), and the star-forming gas (green dot-dashed line). The black arrows correspond to the epochs during which the stellar mass grows by more than 50\% in 5~Myr. For comparison, we also report the stellar half-mass radius as a blue dotted line. While the neutral gas size increases with time, reaching several kpc, the stellar component increases at a much slower pace, consistent with the average trend observed for the star-forming gas. The strong fluctuations in the radius of the star-forming gas are caused by the alternating phases of gas inflows/SF bursts and stellar feedback evacuating the cold dense gas. Moreover, the gaseous star-forming region evolves coherently with the H$_2$-dominated region, in agreement with the expectations that most of the stars form in molecular clouds, but also that not all molecular hydrogen is necessarily star-forming.}
\label{fig:size_z}
\end{figure}

In Fig.~\ref{fig:size_z}, we show the evolution of the half-mass radius for the different gas phases and the stellar distribution. The neutral gas is shown as a black solid line, H$_2$ is shown as a red dashed line, and star-forming gas as a green dotted line, and the stellar half-mass radius is reported with a blue dotted line. While neutral gas forming the galaxy disc continuously grows with time, extending up to a few kpc in size, the molecular component remains more concentrated within about 1~kpc.
This is on the low-end side of the size distribution found by \citet{venemans20}, where the typical deconvolved size is about 1--2 kpc.

As expected, despite the slightly stronger fluctuations, the star-forming region within the galaxy consistently follows the molecular gas distribution. This result, that naturally arises in our simulation, is in perfect agreement with \citet{lupi18}, confirming that the assumption of an H$_2$ abundance dependence of the SFR prescription is not necessary. The strong fluctuations observed for H$_2$ and star-forming gas with respect to neutral gas reflect the alternation between phases of strong gas inflows and SF bursts and phases dominated by stellar feedback that evacuate the gas and dissociate H$_2$. In particular, the physical  model employed here suppresses SF in highly unbound, warm gas, consistently with the expectation that stars form in cold, dense clouds which are dispersed after a few Myr due to stellar feedback. This self-regulation results in strong oscillations of the half-mass radius of the star-forming gas. Compared to the extended neutral gas and the more compact molecular and star-forming gas, the stellar component tends to remain even more concentrated, with the half-mass radius extending up to $\sim 300$ pc, in agreement with simulations of quasar hosts \citep{tenneti19,marshall19} and observations of $z=6-8$ galaxies \citep[e.g.][]{shibuya15,kawamata15}, confirming the difficulty in resolving the stellar morphology of high-z galaxies and quasar hosts.

\subsection{Inflowing and outflowing gas in the galaxy}
\begin{figure*}
\includegraphics[width=\textwidth,trim=3cm 0 3cm 0,clip]{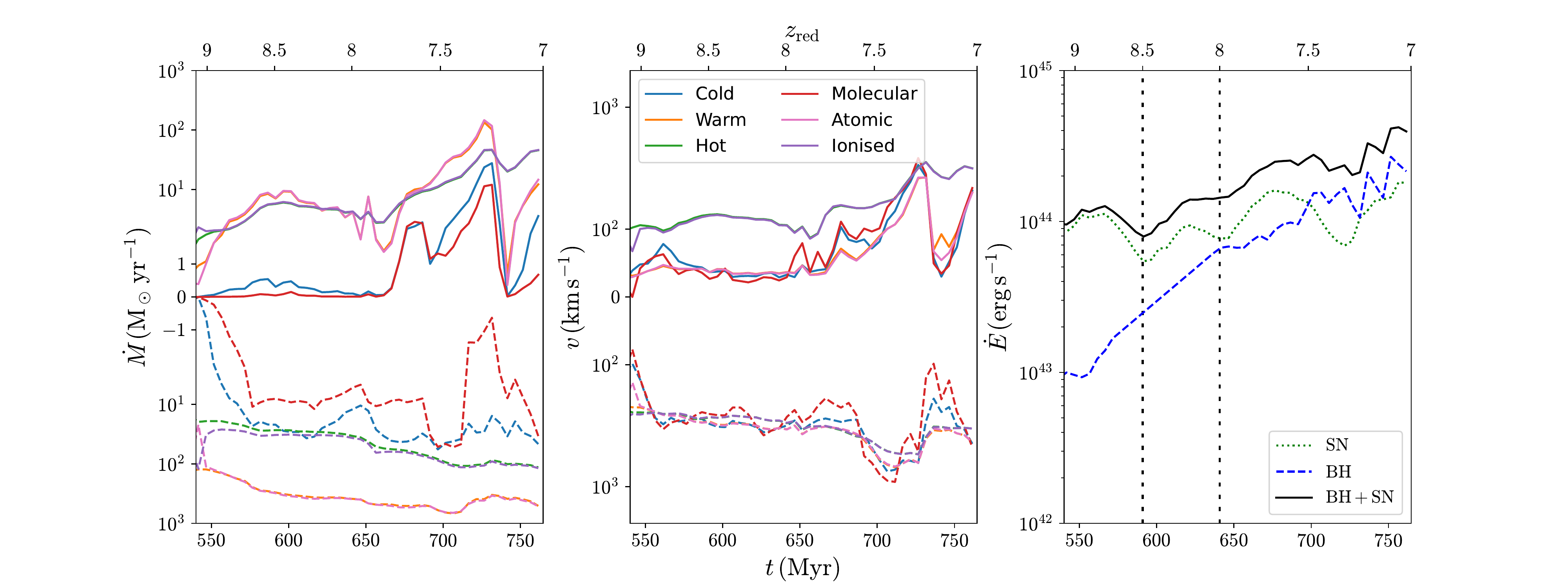}
\caption{Inflow (negative values) and outflow (positive values) rates at $20$~kpc in the redshift range 9--7 for the different phases (left panel), average radial velocities (middle panel), and SN and BH energy deposition rates (right panel).  While the inflows always dominate over outflows, the outflow rate shows a significant increase below $z=8$ (identified in the right panel as a black dashed line), corresponding to the time when the AGN feedback becomes comparable to the total SN feedback injection (as shown in the right panel). If we consider the central 0.2~kpc, the AGN impact becomes dominant much earlier, around $z=8.5$, as shown by the left-most black dashed vertical line in the right panel. The average velocities, on the other hand, only show a moderate increase, suggesting that the AGN is not able to push a lot of gas at the thousands of km/s velocities observed in very fast winds.  The presence of such fast outflows from $z\sim 6$ quasars is debated: in the text we discuss our results in relation to observations and other theoretical investigations.
}
\label{fig:inout}
\end{figure*}

During the cosmic evolution, a lot of gas is accreted onto the galaxy via mergers and smooth accretion from the cosmic web. While part of this gas is turned into stars or accreted onto the central MBH, a significant fraction can be expelled by SNe and AGN feedback, in different phases. Observationally, AGN outflows in the ionised, neutral, and molecular phase have been observed, in some cases even with very large velocities. Here, we assess how the complex interplay between gas accretion and stellar/AGN feedback affect the dynamics of the multiphase ISM, by computing the mass in the different phases (ionised, neutral, and molecular hydrogen) that is inflowing or outflowing with respect to the galaxy.

Furthermore, in most numerical simulations, the three phases are identified solely by applying simple temperature cuts. In our simulation, instead, we have the ability to directly follow the fractional abundance of H$^+$, H, and H$_2$, and this also allows us to compare our accurate identification to the common approach.

For this analysis, we filter the gas cells exhibiting a negative or positive radial velocity that fall within a spherical shell located at 20~kpc from the galaxy centre (with size $\pm 6$~kpc), that corresponds to the distance at which molecular outflows have been observed by \citet{maiolino12}.  For the temperature cuts, we assume that $T>3\times 10^4$~K corresponds to ionised (hot) gas, $300\,{\rm K}<T\leq3\times 10^4$~K to neutral (warm) gas, and $T<300$~K to molecular (cold). For the chemical abundance definition, instead, we select particles where at least 50 per cent of the  mass is in H$^+$ (ion), H (neutral), or H$_2$ (mol).

The results are reported in Fig.~\ref{fig:inout} in the redshift range 9--7, as above. In the left panel, we show the inflow (outflow) rate in the different phases as dashed (solid) lines, with the same colours/markers corresponding to the same phase. As the galaxy evolves, the inflow rate almost steadily increases in all phases, due to the continuous accretion from the cosmic web and via mergers, despite the concurring outflow phases.

When comparing the common temperature cuts with the chemistry-based ones, we notice that (i) most of the inflowing/outflowing mass is in the atomic phase, that is composed by warm gas only (suggesting that our cut at $3\times 10^4$~K is a good choice to identify the mostly neutral phase). 
This is also evident from the mass of ionised gas, that well overlaps with the hot component.
Molecular gas and the cold phase are instead not always overlapping, despite being very similar (within a factor of 2), and this suggests that 300~K represents a good but not perfect threshold to securely identify the mostly molecular regime. This is somewhat expected, since the neutral/molecular transition is much more sensitive to the non-equilibrium conditions and the non-ionising radiation flux than the neutral/ionised one. 

In the middle panel, we report the mean inflow (outflow) radial velocity of the gas in the different phases. Although the mass rates vary significantly among the different phases, their inflow (outflow) velocities are almost identical, with only slightly larger fluctuations for the molecular phase, resulting from its extremely low mass rate that makes it more sensitive to variations in the gas dynamics in the galaxy.

Finally, in the right panel, we show the SN and BH energy deposition rate as green dotted and blue dashed lines, respectively, and their sum as a black solid line. The SN luminosity has been estimated from the average SFR in the galaxy (see Paper I), assuming discrete SF bursts lasting 5~Myr each (corresponding to the time separation between the outputs), and estimating the number of SN exploding according to the stellar lifetimes of a Chabrier IMF stellar population \citep[see][for details]{lupi19b}.
For each SN, we assumed a released energy of $10^{51}$~erg. On the other hand, for the AGN, we estimated the luminosity from the average BH accretion rate (Eddington-limited), assuming a radiative efficiency $\eta_{\rm rad}=0.1$ and a coupling efficiency $\eta_{\rm fbk}=5\times 10^{-3}$. The black dashed vertical lines correspond to the times at which the impact of AGN feedback becomes comparable to that of SNe within 0.2~kpc (left-most line) and 2~kpc (right-most line), respectively.

\begin{figure*}
\includegraphics[width=\textwidth,trim=2.8cm 1cm 3.5cm 0,clip]{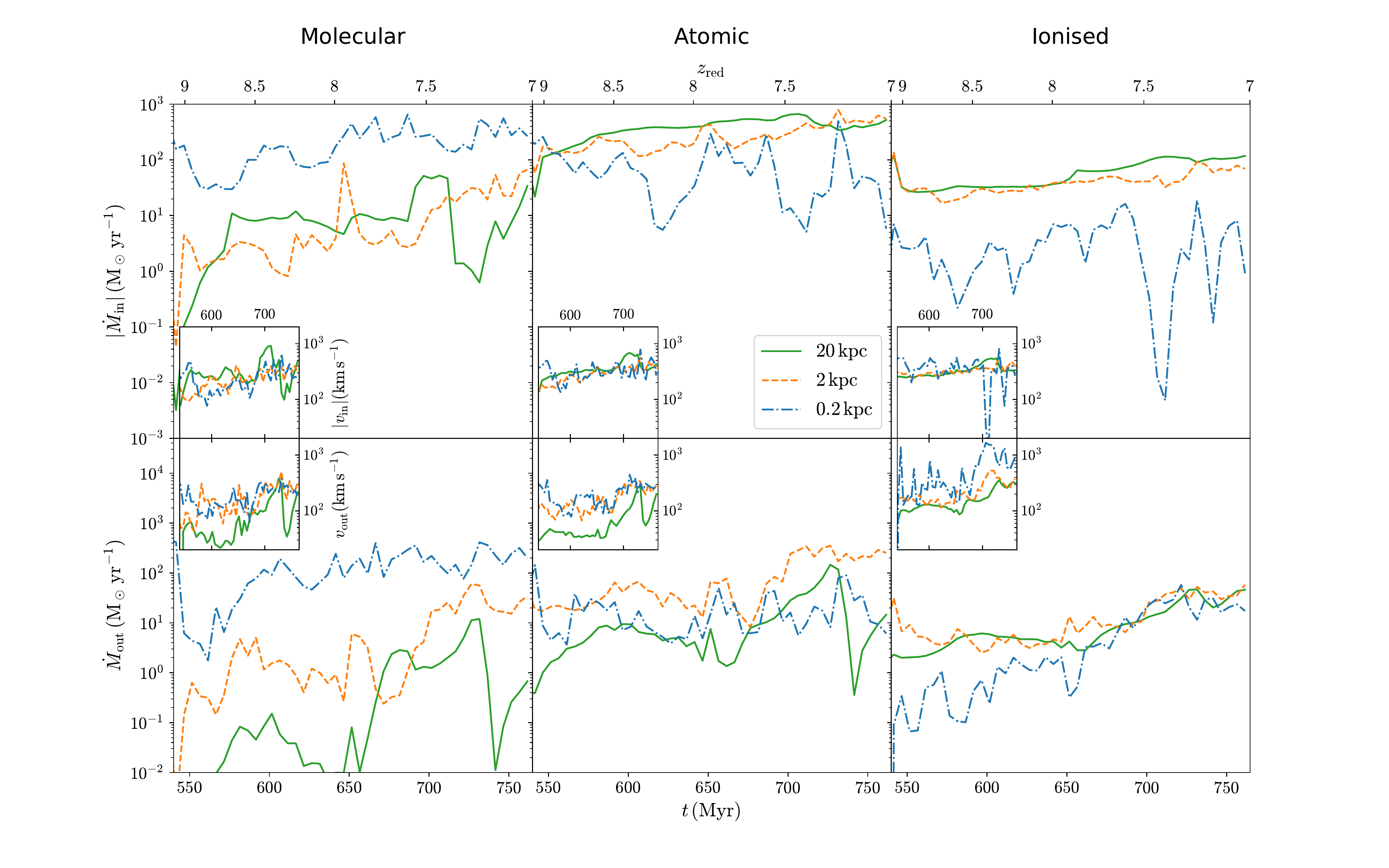}
\includegraphics[width=\textwidth,trim=2.8cm 0.2cm 3.5cm 1.5cm,clip]{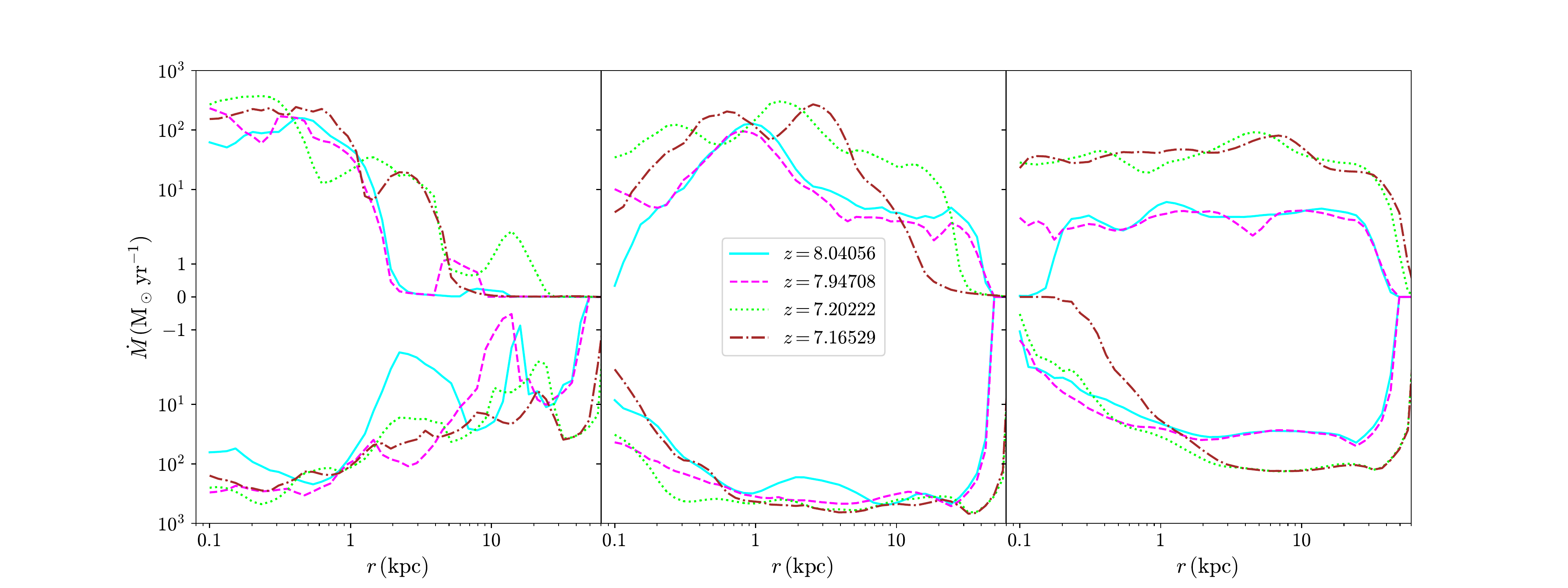}
\caption{Inflow (first row) and outflow (second row) rates in the redshift range 9--7 for the molecular (left panels), atomic (middle panels), and ionised gas phases (right panels), with the corresponding average radial velocities shown in the inlets (middle panel). The green solid lines correspond at a spherical surface at $20$~kpc, the orange dashed ones at 2~kpc, and the blue dot-dashed at $0.2$~kpc. In the bottom panels, instead, we report the inflow (negative) and outflow (positive) rate radial profiles around the time of disc settling ($z\sim 8$) and the time of an AGN feedback burst ($z\sim 7.2$). While, at large scales, inflows mainly occur in the neutral/ionised phase, with small amounts of molecular gas ($\lesssim 10\%$), colder and denser gas is present on smaller scales, with properties typical of the ISM, resulting in the inflows being dominated by the molecular component, as can be seen from the top and bottom panels. Inflow velocities are also quite similar, especially for the neutral/ionised components, because of their relation with the large scale accretion onto the halo and the global instabilities in the galaxy. As for the outflows, a clear trend can be noticed, where most of the outflow mass is molecular near the galaxy centre (up to $\sim 1$~kpc), and evolves to the neutral (up to a few kpc in size) and ionised phase ad the gas moves outwards. Notice also that the outflow rate generally decreases at larger scales, being partially stopped by the interaction with the surrounding medium. Especially at lower redshift, the outflow velocities exceed a few hundred km~s$^{-1}$ at small scales, but significantly decrease at larger scales, in all the phases.}
\label{fig:inout_comp}
\end{figure*}

By comparing the inflow/outflow evolution with the SN and BH energy injection rates, we see that, for $z\gtrsim 8$, the outflow rate is mainly SN-driven, and is on average one order of magnitude smaller than the inflow. Below $z=8$ instead, the central BH contribution to the energy injection starts to become important, and  exceeds that by SNe below $z\sim 7.5$. $z\sim 8$ also marks the transition to a more AGN-dominated driving of the outflows, with the outflow rate quickly increasing up to a significant fraction of the net inflow rate. This change can be also observed in the radial velocity of the outflowing gas, that increases by a factor of 2--3 below $z=8$, unlike that of the inflowing gas, that varies only mildly. Although noticeable, such a velocity change suggests that the AGN is not able to eject gas at thousands of km s$^{-1}$. 

Similar analyses of the outflows in high-redshift quasar hosts have been performed in previous works, although at completely different resolutions and with different sub-grid models \citep{costa15,barai18,ni2018}. We also recall that the final MBH mass is $\sim 6\times 10^8\msun$. With respect to \citet{costa15}, in which they analysed the outflow properties in a halo similar to ours at $z= 6.6$, our galaxy at $z\leq 8$ shows a few times less massive outflows (identified as the gas mass above 300~km~s$^{-1}$), although the MBH mass is similar ($\sim 8\times 10^8\msun$). A similar result is also obtained when we apply the criteria in \citet[][with MBH masses between $4\times 10^8\msun$ and $4\times 10^9\msun$]{barai18} and \citet[][MBH mass $\sim 8\times 10^8\msun$]{ni2018}, where our outflows are generally lower. The only exception is at $z=7.3$, when the criteria in \citet{barai18} yield a mass outflow rate in our run in broad agreement with the observational results by \citet{cicone15}, similar to what found in \citet{barai18} at lower redshift. Although the weaker outflows in our runs could in principle be associated to a somewhat less effective AGN feedback, in Appendix A of paper I we showed that our feedback efficiency is expected to result in a more effective heating of the gas around the MBH than, for example, that in BlueTides \citep{dimatteo17}. In the light of this, there are several additional effects that should be taken into account when comparing our results with other works, i.e. our galaxy at $z\sim 7$ is (within 300 pc) a factor of 2 more massive than in \citet{costa15} and \citet{barai18}, and the corresponding potential well deeper, and our MBH is moderately less massive ($\sim 80$ per cent) and accreting at a lower accretion rate (about 0.5--0.6 times the Eddington limit). Compared to \citet{ni2018}, we should also note that their halo mass is also a factor of two lower than ours, i.e. the escape velocity is lower than in our case. Finally, another important difference is the spatial/mass resolution in the simulations, that can significantly affect how the AGN feedback couples with the gas, i.e. i) a lower resolution results in feedback being coupled at larger scales/lower densities, resulting in an easier escape from the galaxy because of a less resolved interaction with the interstellar gas and a subsequent lower impact of radiative cooling, and ii)  energy is distributed over a much larger mass, potentially artificially enhancing the amount of mass coupled.

Compared to observations, our outflows appear weaker than those reported by \citet{cicone15} and \citet{bischetti19}, where they identified strong features with radial velocities above $v_{\rm r}\sim 700-900\kms$. However, the presence of these outflows is still debated, with different studies suggesting weak or a total lack of strong outflowing gas features in the spectra \citep{decarli18,novak20}. In particular, in \citet{novak20}, the authors also compared their stacked spectra with the MassiveFIRE simulation suite, where only SN-driven outflows were included, and they found good agreement between the [CII]/dust continuum observed profiles and the simulated dust profiles, reinforcing the idea that these outflows are not so common. 

Moreover, in the case of AGN feedback, the distance at which the outflow is measured plays an important role. To assess how the inflowing/outflowing gas evolves in the quasar host, in Fig.~\ref{fig:inout_comp} we compare the inflow (first row)/outflow (second row) properties at three different distances from the centre of the galaxy, i.e., 0.2~kpc (blue dot-dashed lines), 2~kpc (orange dashed lines), and 20~kpc (green solid lines) in the different phases, i.e. the molecular phase (left panels), the neutral one (middle panels), and the ionised one (right panels). In addition, to better highlight the spatial distribution, we show in the bottom panels the inflow (negative) and outflow (positive) rate radial profiles around the time of disc settling ($z\sim 8$, solid cyan and dashed magenta lines) and that of an AGN feedback burst ($z\sim 7.2$, dotted lime and dash-dotted brown lines). Inflow velocities, resulting from the large scale accretion and the global gravitational instabilities in the galaxy, behave in a very similar way in all the phases. The inflow rate instead shows a clear dependence on the distance, with most of the gas at large scales neutral or ionised (notice the clear drop of the ionised phase around a few kpc, while molecular gas is mostly present near the centre, coming from the ISM of the galaxy. 

For the outflows, the picture is different, with velocities easily exceeding hundreds of km~s$^{-1}$ near the centre, but get significantly smaller as we move to 20~kpc from the galaxy, because of the interaction with the surrounding circum-galactic medium and the gain of potential energy while escaping from the halo. The outflow rates clearly show that most of the ejected mass is in molecular form near the galaxy (up to about 1~kpc), which is then dissociated/ionised as the gas moves outwards, shocks with the surrounding gas, expands, and becomes less shielded from radiation. The radial profiles clearly support this picture, with the peaks in the outflow rates shifting to larger radii depending on the phase, from the molecular to the ionised one respectively. Interestingly, the steep rise in the molecular outflow rate near the AGN (left panels) occurs much earlier than in the ionised phase (right panels), with the neutral gas not showing any clear increase with redshift. Around $z=8.5$, the AGN feedback starts to dominate over SN feedback in the centre (see the leftmost dashed vertical line in Fig.~\ref{fig:inout}), but the thermal energy injected is not sufficient to ionise a lot of mass (as reflected by the small bump in the ionised phase), resulting in a rapid increase of the molecular outflows only. At $z\lesssim 8$ instead, the thermal energy released by the AGN becomes large enough to quickly ionise the surrounding gas, pushing it outwards where it interacts with the neutral and molecular gas clumps lifted by SN feedback, producing a net rise of the outflow rate in both phases at larger radii (2 and 20~kpc). 

\begin{figure*}
\includegraphics[width=\textwidth,trim=0cm 0cm 2cm 0cm,clip]{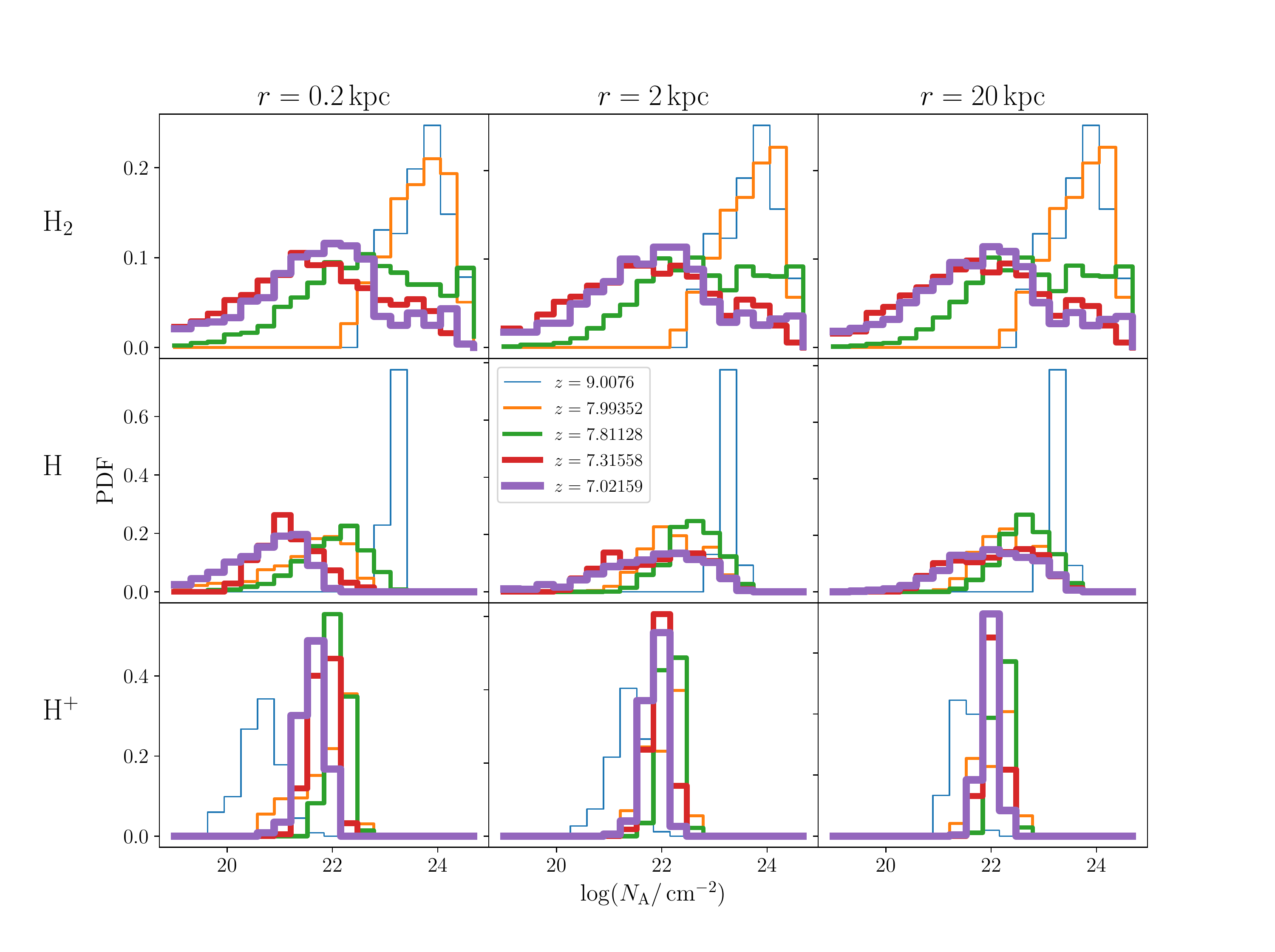}
\caption{Column density distribution of different gas phases at different redshift/times up to 200~pc (left panels), 2~kpc (central panels), and 20~kpc (right panels). From top to bottom, we show H$_2$, H, and H$^+$, with the redshift evolution identified by the different colours (blue, orange, green, red, and purple) and line sizes (from thick to thin). While H$_2$ and H evolve with redshift, spreading towards lower column densities, consistent with the settling of the galactic disc, H$^+$ remains peaked around $N\sim 10^{22}\rm\, cm^{-2}$, hinting for a more uniform distribution on different lines of sight.  
}
\label{fig:NPDF}
\end{figure*}

To further confirm this idea, we focussed on $z=7.8$ and $z=7.3$, that correspond to the epochs at which the cold gas/molecular outflow rate peaks at $r=20$~kpc, and traced forward in time the particles belonging to the outflow, estimating  the total abundance of H$_2$, H, and H$^+$ as a function of time. At early times, we observed indeed a rapid conversion of about half of the molecular gas mass into atomic form and a significant increase of the ionised gas component (that remains nevertheless subdominant). At later times, instead, no further conversion occurs, but a small fraction of the atomic gas starts to reform molecular hydrogen, likely because of the shocks occurring with the surrounding gas, that increase the density hence boosting the cooling rate, thus leading to a flattening of the molecular/neutral ratio. This additional analysis further supports the idea that cold gas clumps forming in the galactic disc are initially lifted by SNe first and then pushed by AGN feedback, as discussed in \citet{biernacki18}. In addition, in our study molecular outflows at galactic scales ($r\gtrsim 1$~kpc) are subdominant compared to their neutral counterpart, as they merge with the entrained gas after interaction and dissolve as they move outwards, converting into neutral and then partially ionised gas, because of their decreasing density that makes the gas less shielded from the surrounding radiation field. On top of this, the outflows we find are produced by a MBH of at most $6\times 10^8\msun$ (at $z=7$), which is smaller than those responsible for the strong molecular outflows observed at lower redshift, hence they cannot be directly compared. Nevertheless, while this result seems to disfavour the idea that molecular gas cannot be accelerated and must form out of the thermal instability cooling within the outflow \citep[see][for a detailed investigation]{richings18a,richings18b}, the resolution of our simulation, that is not enough to properly model this process, and the low mass outflow rate we observe at large scales, do not allow us to give a definitive explanation of the origin of the observed molecular outflows, but only a hint of it.

\subsection{AGN obscuration}

As soon as the impact of AGN feedback onto its host becomes relevant, gas is pushed away, clearing channels through which the radiation emitted by the accreting MBH can be directly observed. However, in many cases (especially at high redshift when gas is continuously flowing onto galaxies from large-scale filaments) the gas density around the MBH \citep[and in the galaxy, see][]{circosta19} is so high that the AGN remains obscured, hindering detection in optical-UV bands and sometimes also in X-rays. 
Because of this limitation, the optical-UV identification of high-redshift quasars in surveys like SDSS is significantly biased towards the most luminous and energetic MBHs, and may potentially miss an important fraction of high-redshift quasars, in the case most of them are highly obscured. It is therefore of utmost importance to assess the typical obscuration of high-redshift quasars at different stages of their evolution, and how different gas phases contributed to obscuration, to infer the number of potentially missed objects in large observational campaigns and the possible biases in the properties of the detected systems. 

To this aim, we analyse here the covering fraction of different gas phases at different distances from the central AGN as a function of redshift.
Using \textsc{HEALPix}\footnote{\url{https://healpix.sourceforge.io/}}, we cast 972 rays from the BH position, and integrate the density of different gas phases in 1000 logarithmic steps from 5~pc up to $r_{\rm out}=0.2, 2,$~and~20~kpc, as 
\begin{equation}
    N_{\rm A} = \frac{1}{m_{\rm H}}\int_0^{r_{\rm out}}\rho_{\rm A}(r'){\rm d}r',
\end{equation}
where $\rho_{\rm A}(r')$ is the gas species $A$ density along the ray,\footnote{In order to compute the gas density along the ray, we apply a scatter kernel average approach, obtaining $\rho_{\rm A}(r') = \sum x_{{\rm A},j} m_jW(|\mathbf{r'-r_j}|,h_j)$, with $x_{{\rm A},j}$ is the $j$-th particle mass fraction for species A (either HI or H$_2$), $m_j$ the total particle mass, $\mathbf{r}_j$ the position, and $h_j$ the kernel size. $W$ is the smoothing kernel function employed for hydrodynamics, in this case a cubic spline encompassing an effective number of neighbours $N_{\rm ngb}=32$. We also validated this approach by exploring different radial integration steps and number of rays, finding very small variations.}. We therefore only include gas from about 5~pc from the MBH, and we do not treat obscuring gas at smaller scales (e.g., a torus) that we do not resolve. The calculated column density is therefore a lower limit to the actual column density.

The normalised column density distribution for each gas phase  is reported in Fig.~\ref{fig:NPDF}, with H$_2$, H, and H$^+$ from top to bottom. From left to right, the integration is done up to 0.2, 2, and 20~kpc respectively. As expected, the H$_2$ distribution does not change significantly with the maximum distance, since H$_2$ is abundant in the galactic disc, within the central 2~kpc. On the other hand, a drift towards moderately higher column densities is observed for H and H$^+$, consistent with their less concentrated distribution, suggesting that a significant part of the obscuration from the neutral gas comes from the larger scales well beyond the galactic disc. Although not relevant for the obscuration, the peaked distribution of H$^+$ suggests that, unlike other phases, the ionised gas is also more homogeneously distributed over all lines of sight, and never exceeds a few $10^{22}\rm\, cm^{-2}$. With redshift, both H$_2$ and H distributions extend from a single high peak at $z\sim 9$ ($N_{\rm H_2}\sim 10^{24}\rm\, cm^{-2}$ and $N_{\rm H}\sim 10^{23.5}\rm\, cm^{-2}$), when the galaxy disc has not formed yet and dense gas can be found along all lines of sight, down to $z=7$ when most of the dense gas has settled in a thin disc and the AGN feedback has cleared out its surrounding. At $z=7$, the H$_2$ peak can indeed be found around $N_{\rm H_2}\sim 10^{22.5}\rm\, cm^{-2}$, whereas the neutral gas exhibits a roughly bi-modal distribution with two peaks at $10^{21.5}\rm\, cm^{-2}$ and $10^{22.3}\rm\, cm^{-2}$ respectively. 

\begin{figure*}
\includegraphics[width=\textwidth,trim=2cm 0cm 2cm 0cm,clip]{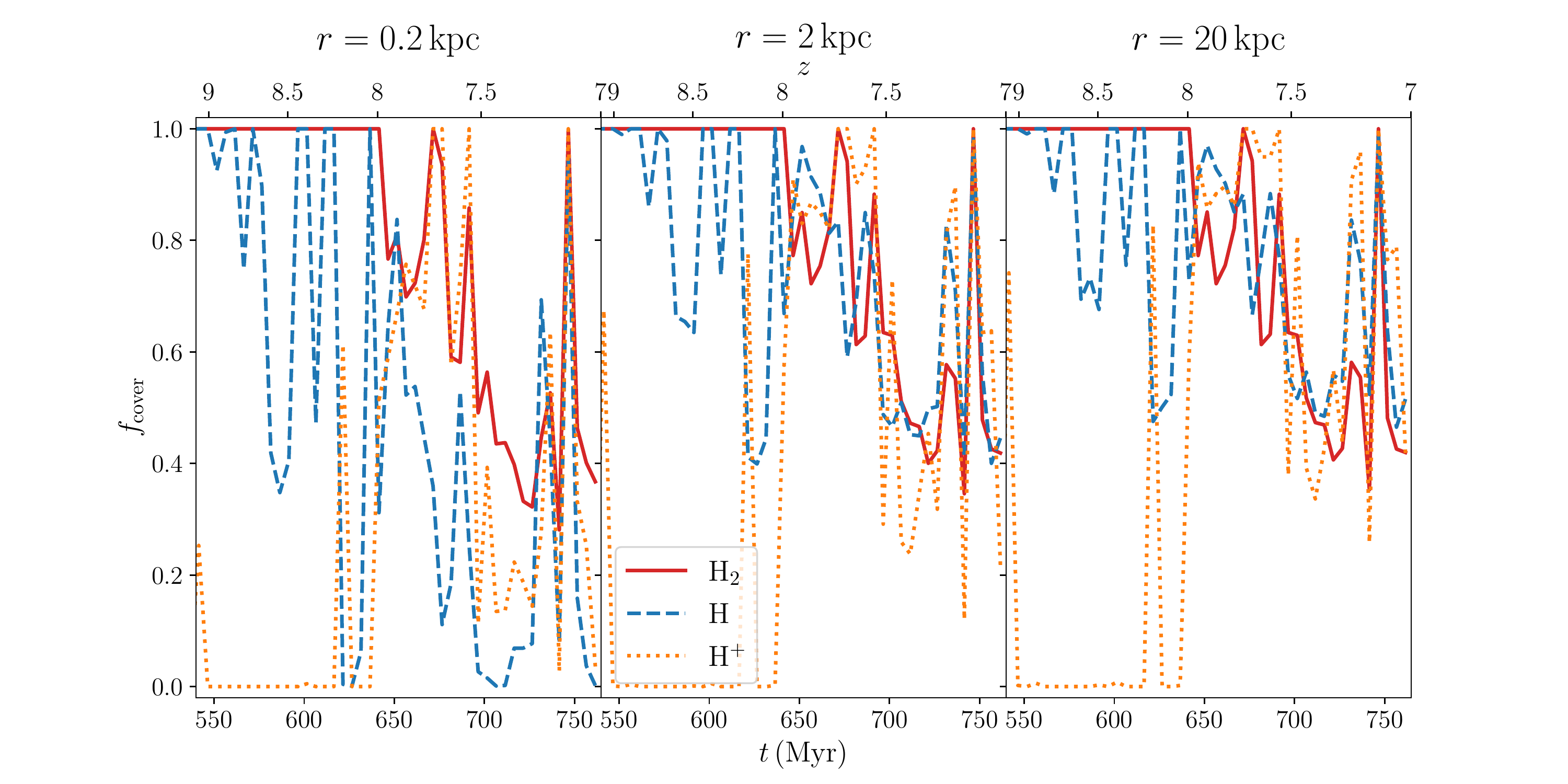}
\caption{Covering fraction of different gas phases as a function of time up to 200~pc (left panel), 2~kpc (middle panel), and 20~kpc (right panel). H$_2$ is shown as a red solid line, H as a blue dashed one, and H$^+$ (which is included for completeness, although it does not contribute to obscuration) as an orange dotted one. H$_2$ contributes the most to the actual obscuration, with the largest effect being at small scales/higher densities, as expected. Below $z=8$, the disc settling starts reducing the obscuration to  about 70-80\%, that finally drops to 40-50\% by $z=7$ after AGN feedback has cleared out its surroundings.  On the other hand, neutral hydrogen dominates at larger scales where the density is lower and the gas is more sparse. This results in strong fluctuations with time, which depend on the evolving location of gas clouds along different lines of sight. Finally, H$^+$ is always subdominant at high redshift, and becomes more important only after the AGN has started to heat up and ionise gas in its surroundings. 
}
\label{fig:fcover}
\end{figure*}

To better assess how the different distribution and evolution affect the actual obscuration, in Fig.~\ref{fig:fcover} we show the time evolution of the covering fraction $f_{\rm cover}$ of different gas phases in the quasar host up to 200~pc (left panel), 2~kpc (middle panel), and 20~kpc (right panel), where $f_{\rm cover}$ is the number fraction of lines of sight with $N_{\rm A}>10^{22}\rm\, cm^{-2}$. The molecular phase is shown as red solid lines, the atomic one as blue dashed lines, and the ionised one as orange dotted lines. Notice that $\rm H^+$ does not actively contribute to the obscuration (it does not absorb photons), but is included anyway for comparison. 

At higher redshift ($z>8$), the BH is heavily obscured already at 200~pc, with most of the obscuration coming from molecular gas, ubiquitously presents in its surrounding (this is also consistent with the continuous accretion at the Eddington limit). Neutral gas becomes important only at larger scales, well beyond the extent of the galaxy host, and exhibits strong fluctuations due to the rapid variations of the column density distribution, affected by gas inflows and stellar feedback. At lower redshift, the disc settling and the clearing of the central region by the AGN feedback result in a decrease of the obscuration to about 0.4-0.5, with molecular gas still dominating the central regions and neutral hydrogen contributing mainly at larger scales. Interestingly, in this phase the fluctuations in the covering fraction of both H$_2$ and H become more similar, especially outside the central kpc, and this is consistent with the disc settling, which results in most of the neutral and molecular gas staying near the disc plane, except for stellar and AGN feedback events, and the rest of the sky remaining clear from gas, hence not obscuring. Ionised gas, which is in any case unable to provide obscuration, is mainly distributed at large scales beyond the extent of the disc, but always in a limited amount, especially at higher redshift, when the covering fraction is almost zero at all scales. 

Similar evolutionary trends have been found also by \citet{ni20} analysing a sample of quasars from the BlueTides simulation, in which the total hydrogen column density PDF becomes larger with time, and drifts from a highly obscured regime at higher redshifts, when the densities are higher, the galaxy more compact, and the AGN has not been able to clear out its surrounding yet, to moderately lower obscuration. However, due to the lower resolution of their simulation and the inability to properly follow the multiphase ISM did not allow them to disentangle the contribution of the different phases, as instead we do here.  \citet{trebitsch19} analyze a simulation with resolution intermediate between that of \citet{ni20} and ours, and focus specifically on the interplay between AGN feedback and accretion in modulating the column density around the MBH in a high redshift AGN. In galaxy merger simulations, \citet{hopkins06} found generally higher column densities for more luminous AGN, consistent with both our local obscuration measure and that by \citet{trebitsch19}. 

To disentangle the effects on the evolution of obscuration of galaxy-driven changes, namely disc settling, and of AGN-driven changes, namely AGN feedback, we deepen our analysis at two redshifts. The first is around $z=8$, i.e. the time at which the molecular gas column density exhibits a first clear drop in the covering fraction. The second at $z\sim7.2$, when the AGN feedback is expected to dominate over SN feedback. 

\begin{figure*}
{\Large \raggedleft \hspace{0.5cm} Molecular\hspace{3.8cm} Atomic \hspace{3.8cm} Ionised}
\includegraphics[width=\textwidth,trim=0cm 0cm 2cm 2cm,clip]{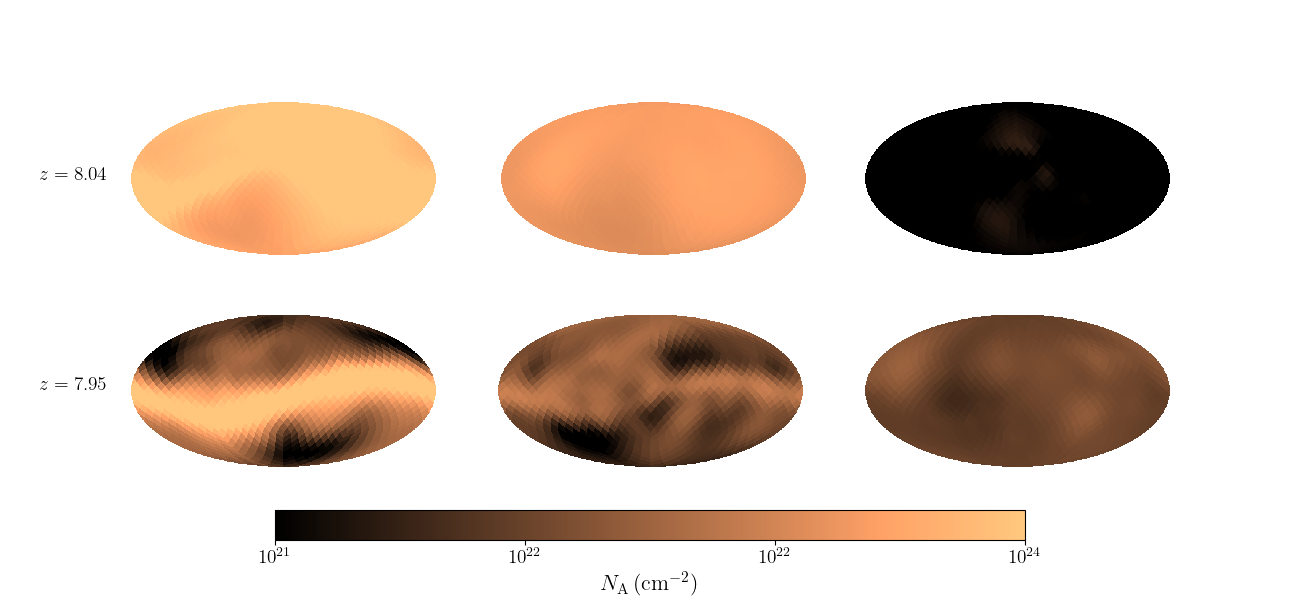}
\includegraphics[width=\textwidth,trim=0cm 0cm 2cm 0cm,clip]{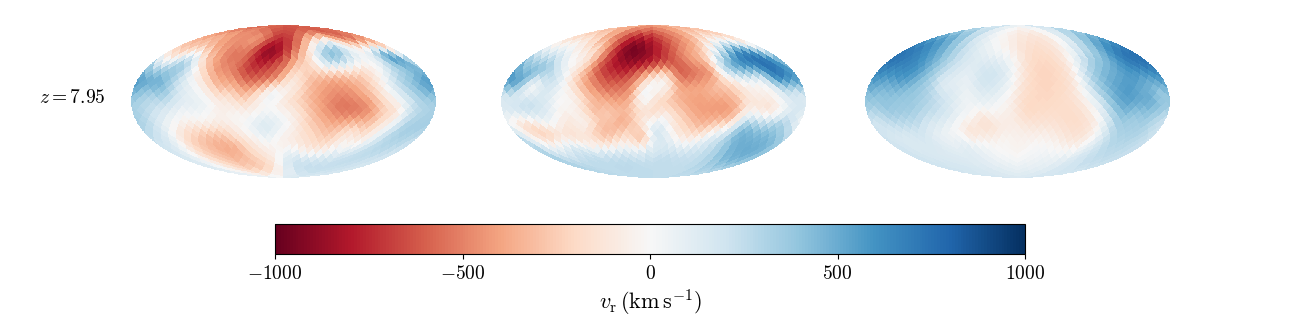}
\caption{Column density of the different gas phases around $z=8$ ($z=8.04$ in the top row and $z=7.95$ in the second one) and average radial velocity (at $z=7.95$ in the bottom one) at 200~pc from the centre. The molecular phase is shown in the left panels, the neutral phase in the middle ones, and the ionised phase in the right ones. Consistently with the redshift evolution already shown in Fig.~\ref{fig:NPDF}, the maps clearly show the formation of the galactic disc, mainly traced by the molecular component, which reduces the covering fraction. The neutral phase exhibits a similar evolution, although some moderately high column densities still persist above and below the disc. Ionised gas instead remains uniformly distributed, and always subdominant in terms of column density relative to the other two phases. The radial velocity distribution shows that the molecular and atomic gas are mostly in inflow, consistently with the formation of the disc.}
\label{fig:Nmap8}
\end{figure*}

We show in Fig.~\ref{fig:Nmap8} the column density map of the different gas phases (molecular on the left, neutral in the middle, and ionised on the right) around $z=8$ (first two rows), along with the average radial velocity at $z=7.95$ (bottom row).. The formation of the disc can be clearly observed in the molecular component, and also partially in the neutral one, whereas the ionised phase remains mostly uniformly distributed. The velocity distribution does not show any clear sign of strong outflows, and is dominated by inflows leading to the settling of the disc, especially in the molecular and atomic phases.

\begin{figure*}
{\Large \raggedleft \hspace{0.5cm} Molecular\hspace{3.8cm} Atomic \hspace{3.8cm} Ionised}
\includegraphics[width=\textwidth,trim=0cm 0cm 2cm 0cm,clip]{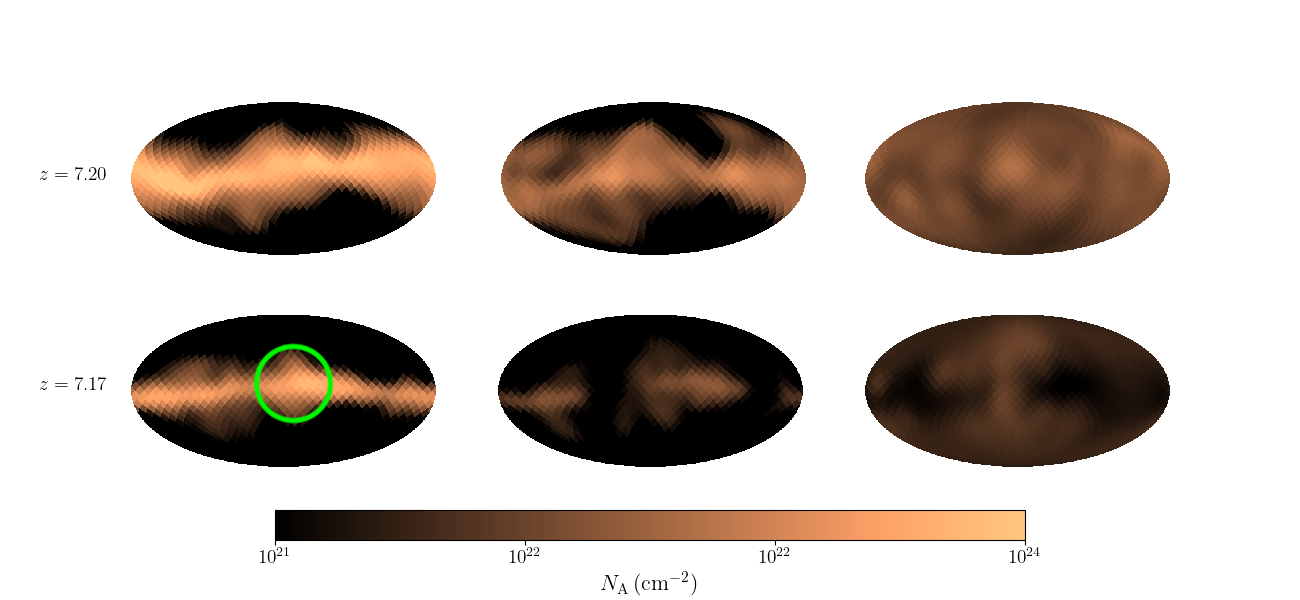}
\includegraphics[width=\textwidth,trim=0cm 0cm 2cm 0cm,clip]{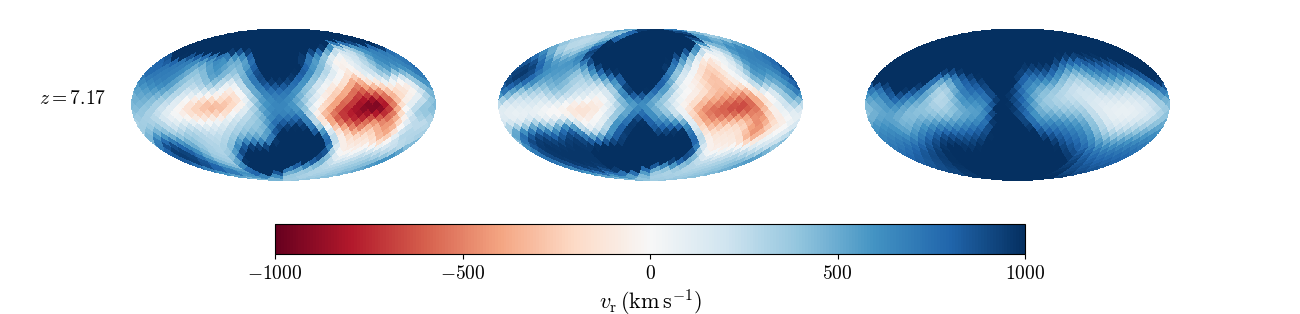}
\caption{Same as Fig.~\ref{fig:Nmap8}, around $z=7.2$ ($z=7.2$ in the top panel and $z=7.17$ in the bottom one). Non axi-symmetric features in the disc can be observed at both times, and inflows and outflows can only be discriminated by measuring the radial velocity. In particular, at $z=7.2$ most of the gas (molecular and neutral) outside the disc plane is flowing onto the galaxy, whereas the almost central gas cloud (highlighted with a green circle) at $z=7.17$ is outflowing at $\sim 500\rm\, km \, s^{-1}$, mainly pushed by the AGN feedback. This can be better seen in the radial velocity distribution (bottom panels), in which the central region exhibits a clear positive velocity extending well above/below the disc in all phases. The effect on the ionised gas is almost negligible, given its very low column density on average.}
\label{fig:Nmap7.2}
\end{figure*}

In Fig.~\ref{fig:Nmap7.2}, we show again the gas distribution in the central 200~pc as in Fig.~\ref{fig:Nmap8}, but now at $z\sim7.2$. Besides the constant presence of the disc, inflows and outflows related to the AGN can modulate the column density. By simply looking at the column density maps, it is hard to assess when the gas is expelled from the galaxy or is flowing onto it, as can be noticed, for instance, comparing the two molecular gas column density projections in the left panels. By further analysing the radial velocities, at $z=7.20$ the features appearing above/below the disc in the molecular and neutral phases are inflowing gas, part of which is likely to cause AGN activity. In fact, shortly after at $z=7.17$ the central overdensity (circled in green in the figure) exhibits a positive outward velocity of $\sim 500\rm\, km\, s^{-1}$, as shown in the radial velocity map in the bottom row, consistent with an outflow effect of the central AGN. Something similar can also be observed in the neutral phase, although the feature is not very prominent. In terms of velocity distribution (bottom panels), inflows are found mainly along the disc, whereas the outflowing gas is mostly seen within a biconical region perpendicular to it in all phases, consistent with the fact that the gas more easily escape through the low density material.

This analysis confirms therefore that the evolution of obscuration comprises two phases. A first one where the column density decreases globally because of disc settling at $z=8$, and this causes a sharp transition in the covering fraction. The second phase is driven by the interplay between inflows and outflows, with fluctuations in the covering fraction, and when the AGN becomes powerful, its feedback greatly contributes to opening lines of sight towards the MBH.

Unfortunately, observations are typically unable to discriminate the role of different gas phases onto obscuration, and simply report the total hydrogen column density. Hence, in order to do a proper comparison, we combine the contribution of the molecular and neutral phase in our simulation and determine the fraction of lines of sight above the critical column density for obscuration, i.e. $N_{\rm H, tot}>10^{22}\rm\, cm^{-2}$ for optical/UV emission and $N_{\rm H, tot}=10^{23}\rm\, cm^{-2}$ for X-rays.
In general, we find that, in the redshift range 7--9, a large fraction, 60 to 90 per cent, of the actual MBH population would be missed in optical-UV surveys, and 30 to 70 per cent even in X-rays, with most of the obscuration coming from the dense gas surrounding the BH.  
In particular, we notice that, on one hand, the high covering fraction at $z\geq 8$, that corresponds to a phase in which the MBH is accreting a lot, but its feedback is not powerful enough yet to clear out its surrounding, would prevent us from identifying such objects, if observed, as quasars. On the other hand, the moment at which AGN feedback starts to expel the gas around $z\sim 7.3$ (see our analysis of the outflows) marks the transition to the actual quasar phase, in which the MBH becomes directly observable, although with a slightly lower accretion rate ($\sim 0.5$ times the Eddington limit; see Fig.~6 in Paper I).

Concluding, our results show that the obscuration in high-redshift quasar hosts is dominated by molecular hydrogen closer to the MBH, with the neutral gas mostly contributing at larger scales. Moreover, the spread PDF of the molecular phase also shows that some lines of sight even reach $N_{\rm H_2}>10^{24}\rm\, cm^{-2}$, i.e. the Compton thick regime, also noted by \citet{trebitsch19}, that highlighted a complex interaction between inflows (increasing the column density and AGN activity) and outflows (decreasing the column density and AGN activity).

\subsubsection{Observational tracers}
\label{sec:obs_emission}
\begin{figure*}
\centering
\includegraphics[width=0.37\columnwidth,trim=1.8cm 5.2cm 1.5cm 6cm,clip]{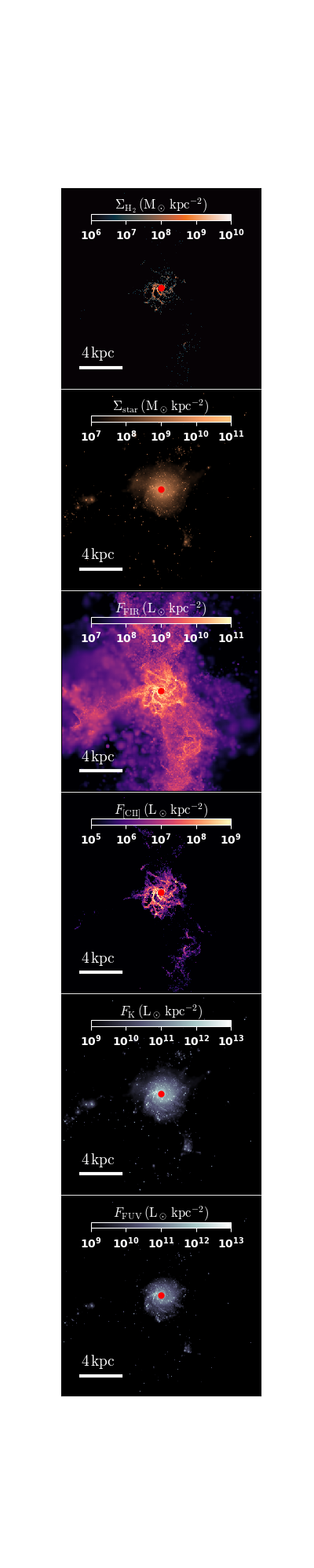}
\includegraphics[width=0.37\columnwidth,trim=1.8cm 5.2cm 1.5cm 6cm,clip]{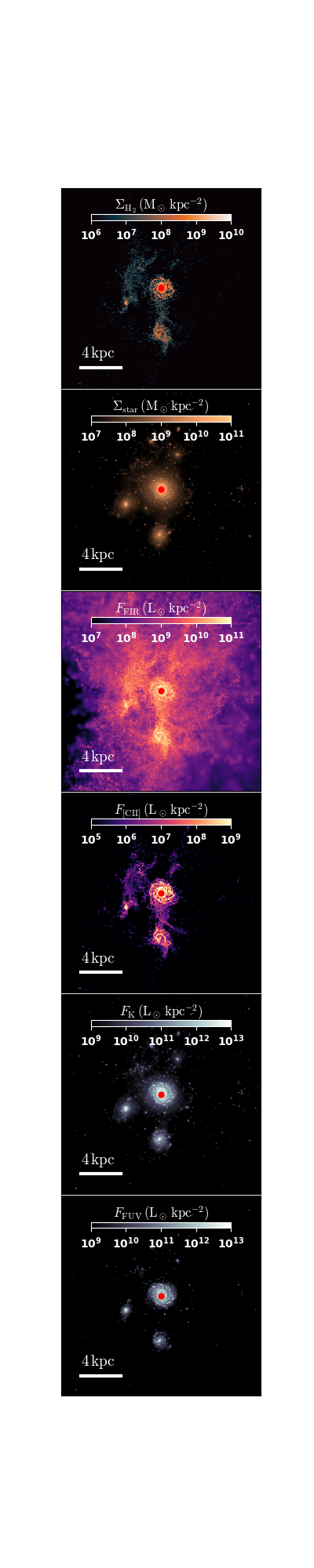}
\includegraphics[width=0.37\columnwidth,trim=1.8cm 5.2cm 1.5cm 6cm,clip]{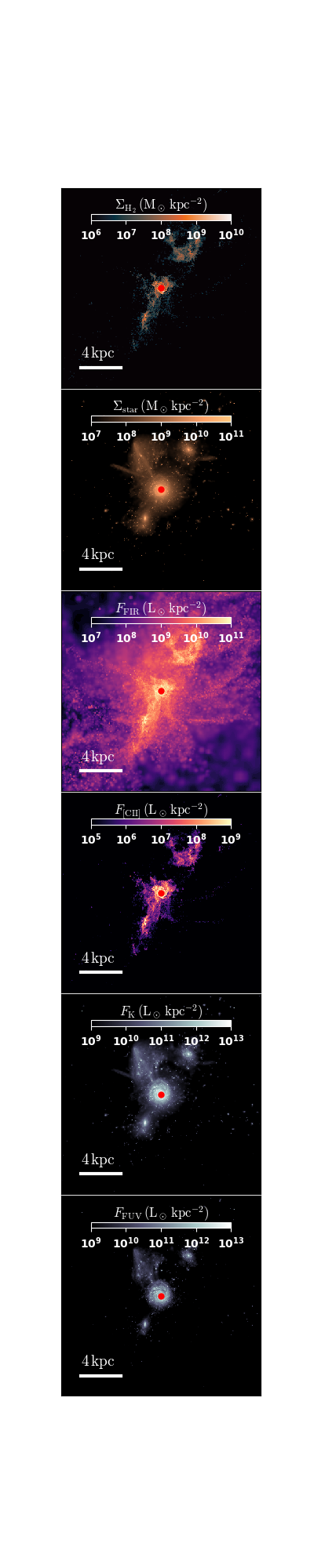}
\includegraphics[width=0.37\columnwidth,trim=1.8cm 5.2cm 1.5cm 6cm,clip]{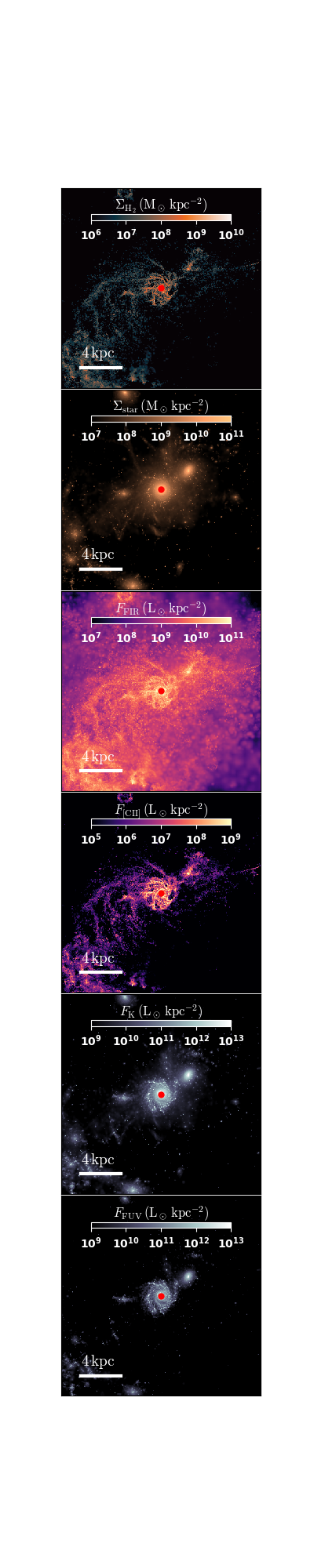}
\includegraphics[width=0.37\columnwidth,trim=1.8cm 5.2cm 1.5cm 6cm,clip]{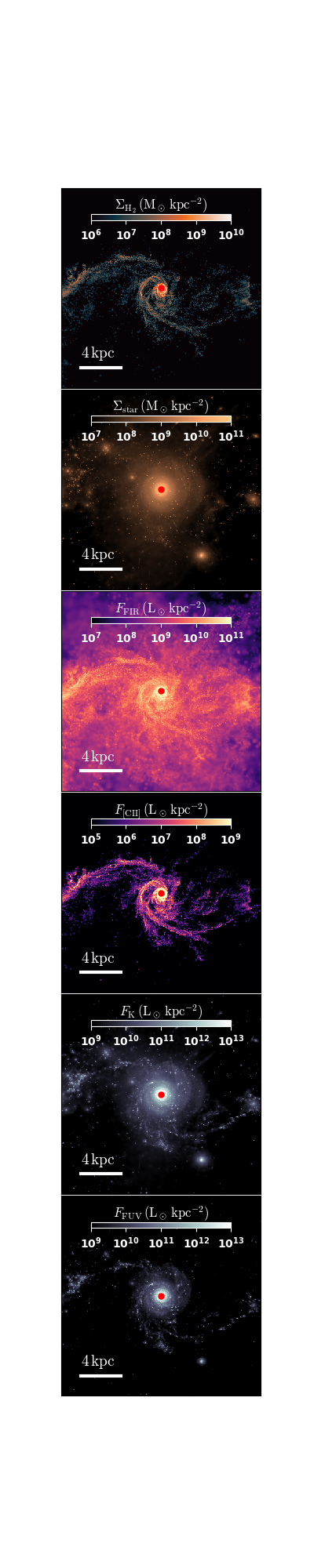}
\caption{Maps of the quasar host at five different redshifts (from left to right: 9, 8, 7.8, 7.3, and~7, corresponding to a time separation $\Delta t = 94, 21, 61,$ and 41~Myr respectively) 
at a resolution of 50~pc, chosen to highlight the morphological evolution of the system. From top to bottom we report the H$_2$ surface density, the stellar mass surface density, the FIR flux, the [CII] line flux, the unattenuated FUV flux and the flux in the rest-frame K band. As seen in the maps, the gas morphology quickly changes because of gas accretion/mergers, and stellar feedback processes over only 200~Myr, whereas the stellar distribution exhibits a smoother evolution.}
\label{fig:tracermaps}
\end{figure*}

\begin{figure*}
\centering
\includegraphics[width=0.37\columnwidth,trim=1.8cm 5.2cm 1.5cm 6cm,clip]{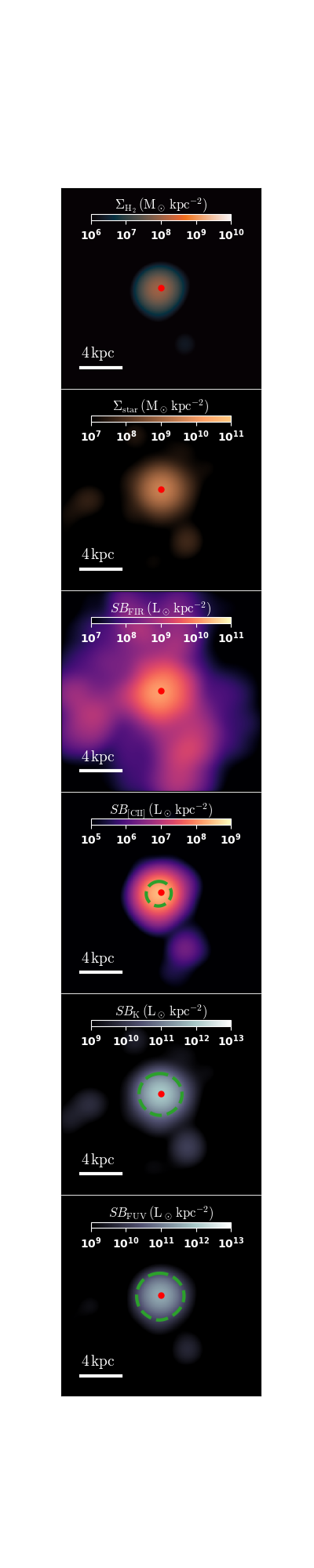}
\includegraphics[width=0.37\columnwidth,trim=1.8cm 5.2cm 1.5cm 6cm,clip]{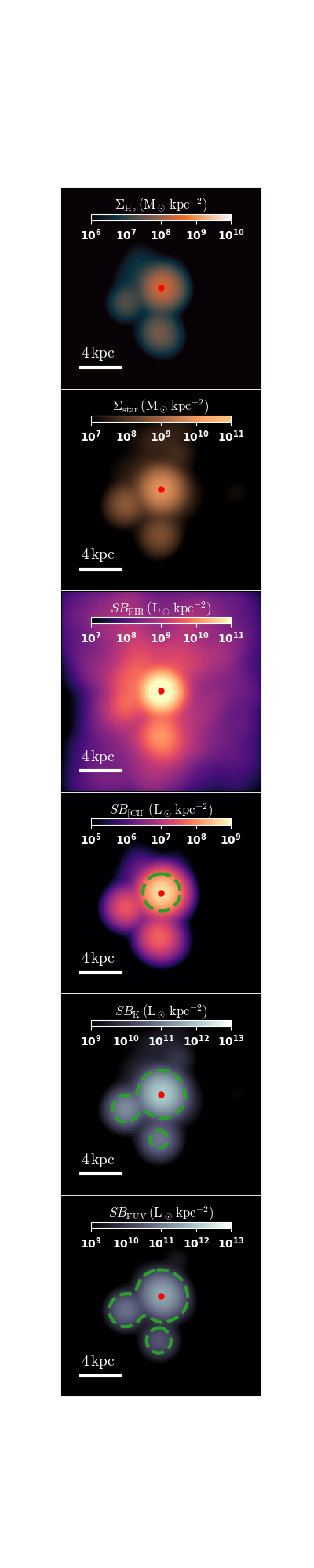}
\includegraphics[width=0.37\columnwidth,trim=1.8cm 5.2cm 1.5cm 6cm,clip]{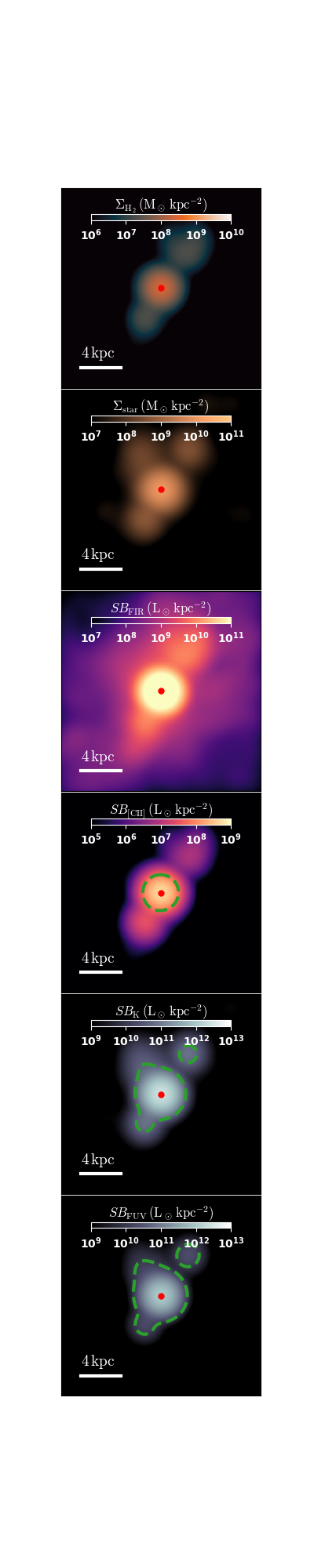}
\includegraphics[width=0.37\columnwidth,trim=1.8cm 5.2cm 1.5cm 6cm,clip]{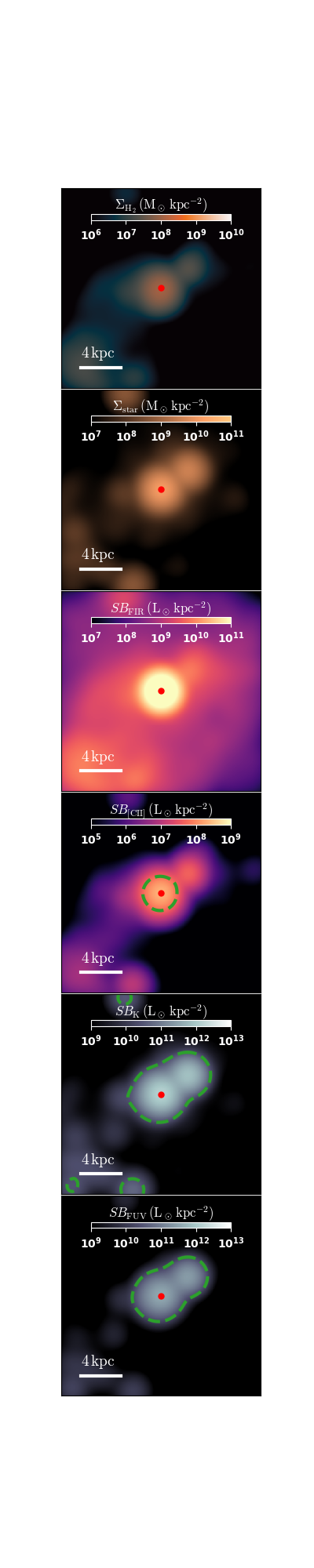}
\includegraphics[width=0.37\columnwidth,trim=1.8cm 5.2cm 1.5cm 6cm,clip]{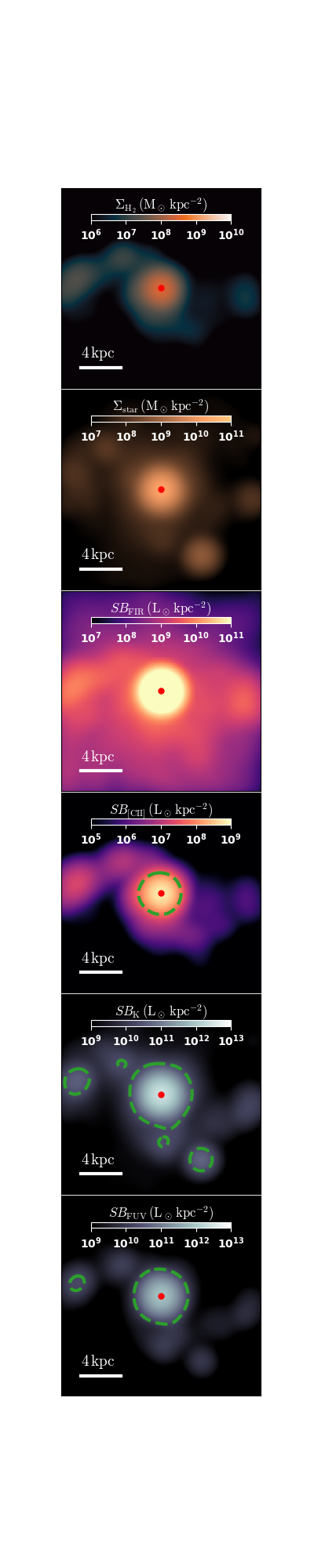}
\caption{Same as Fig.~\ref{fig:tracermaps}, degraded at 2~kpc resolution. All the small-scale features have been smeared out, leaving only the brightest spots still distinguishable from the central galaxy. Nevertheless, depending on the actual luminosity, many of these secondary sources are likely to be confused within the noise, not considered in these maps.}
\label{fig:tracermapslow}
\end{figure*}

\begin{figure*}
\centering
\includegraphics[width=0.37\columnwidth,trim=1.8cm 5.2cm 1.5cm 6cm,clip]{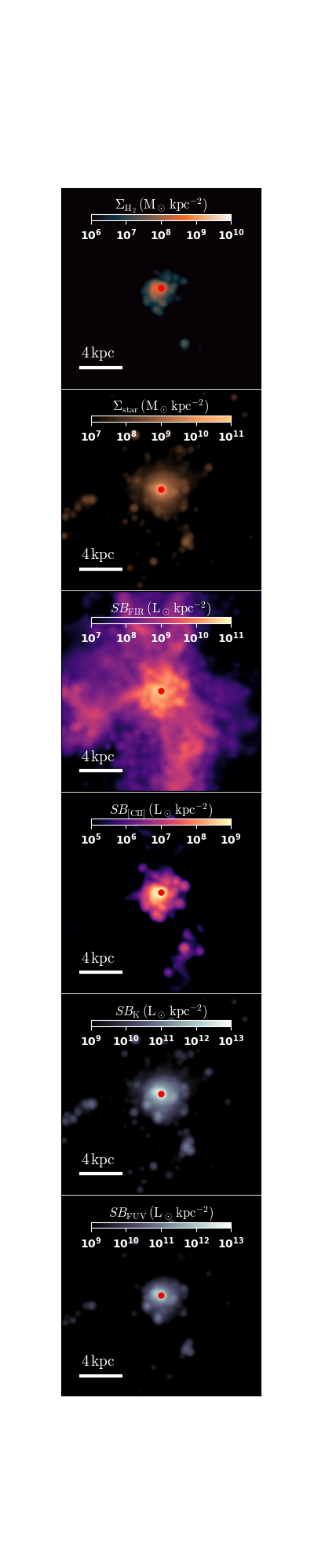}
\includegraphics[width=0.37\columnwidth,trim=1.8cm 5.2cm 1.5cm 6cm,clip]{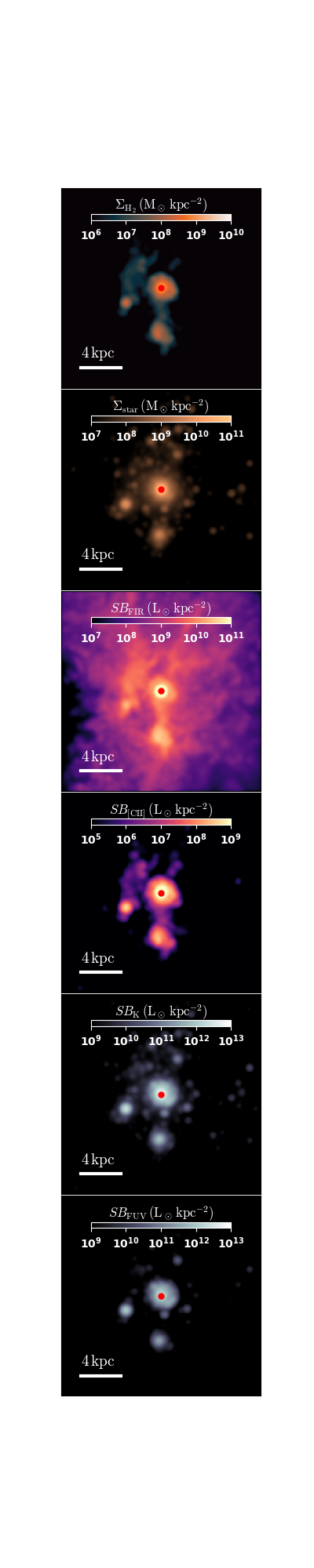}
\includegraphics[width=0.37\columnwidth,trim=1.8cm 5.2cm 1.5cm 6cm,clip]{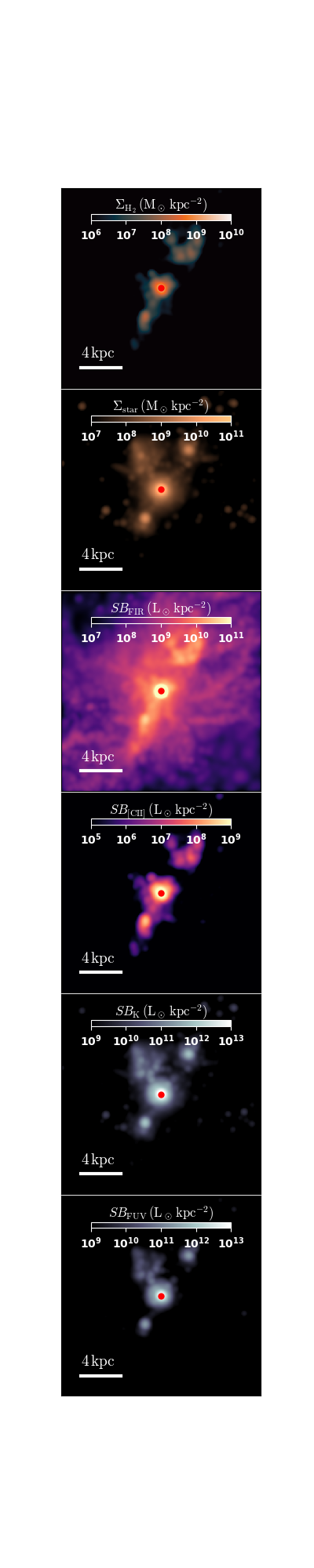}
\includegraphics[width=0.37\columnwidth,trim=1.8cm 5.2cm 1.5cm 6cm,clip]{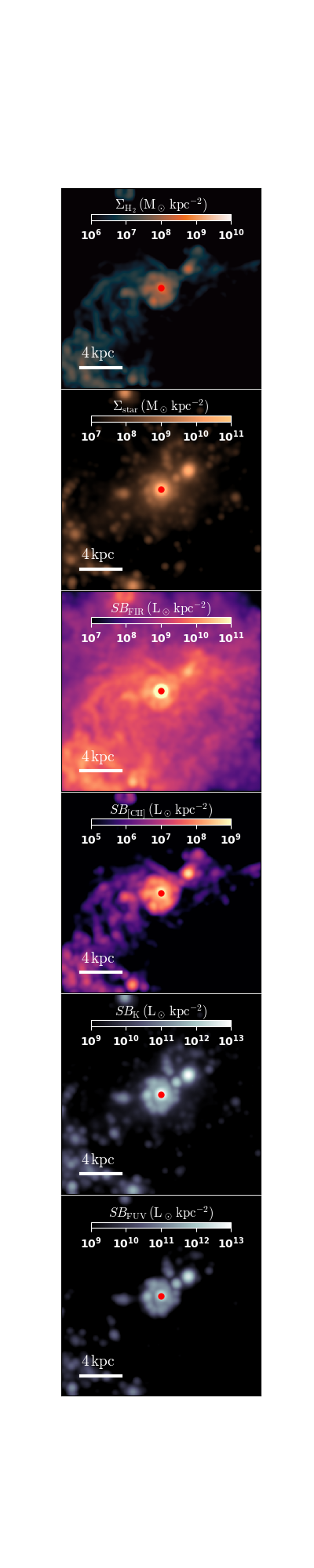}
\includegraphics[width=0.37\columnwidth,trim=1.8cm 5.2cm 1.5cm 6cm,clip]{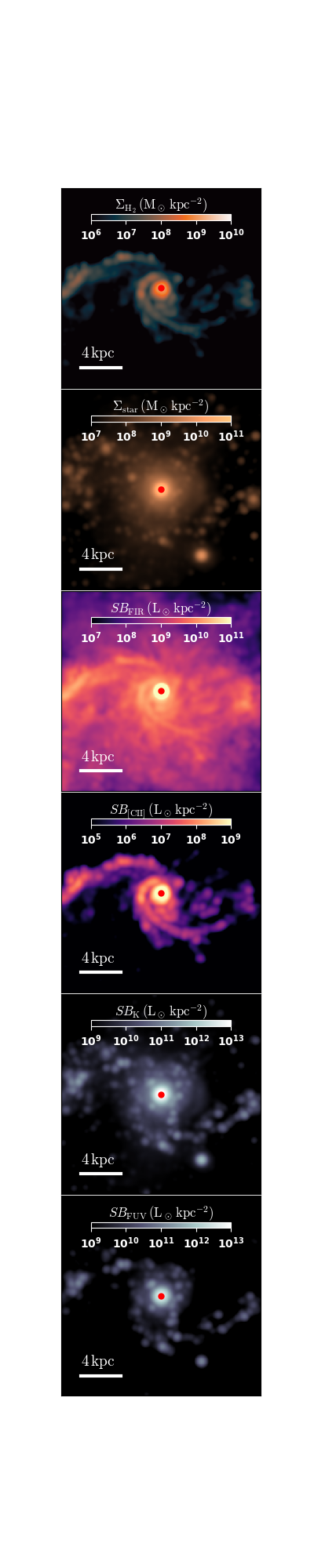}
\caption{Same as Fig.~\ref{fig:tracermaps}, degraded at 500~pc resolution. Unlike for the 2~kpc case, here most of the features are still visible, simply spread over a larger area. Such features may be potentially observable by current and upcoming facilities like ALMA and JWST, provided a long enough integration time.}
\label{fig:tracermapsmid}
\end{figure*}

\begin{table*}
\centering
\caption{Effective size and dynamical mass estimates for different tracers at different redshifts. The first column corresponds to the tracer employed, whereas the subsequent pairs of column correspond to the radius $R$ and dynamical mass $M$ at $z=9,8,7.8,7.3$, and 7 respectively. The first block shows the actual values from the simulation results, whereas the second and the third the observer-like results with a resolution of 200~pc and 2~kpc, respectively. The two masses listed in the observer-like results assume a rotationally-supported disc (first value) or a dispersion-dominated system (second value).}
\begin{tabular}{lcccccccccc}
\hline\hline
\multirow{2}{*}{Tracer} & \multicolumn{2}{c}{$z=$9.0} & \multicolumn{2}{c}{$z=$8.0} & \multicolumn{2}{c}{$z=$7.8} & \multicolumn{2}{c}{$z=$7.3} & \multicolumn{2}{c}{$z=$7.0}\\
&  $R$ (kpc) & $M\, (10^{10}\msun$) & $R$ (kpc) & $M\,(10^{10}\msun$) & $R$ (kpc) & $M\, (10^{10}\msun$) & $R$ (kpc) & $M\, (10^{10}\msun$) & $R$ (kpc) & $M\, (10^{10}\msun$)\\ 
\hline
\multicolumn{11}{c}{\textbf{Simulation (real)}}\\
\hline
Stars && 3.063& &4.936 && 4.328 && 9.611 && 11.200 \\
Gas && 2.181 & &5.301 && 4.305 && 8.228 && 10.640\\
\hline
\multicolumn{11}{c}{\textbf{Resolution: 200 pc ($\sigma_{\rm 2D}=85.1\rm\, pc$)}}\\
\hline
Stars & 0.079 & 1.606/0.769 & 0.090 & 2.168/1.037 &0.086 & 2.174/1.040 & 0.105 &  2.715/1.299 & 0.127 & 3.486/1.668 \\
Gas & 0.145 & 3.530/1.689 & 0.555 & 7.591/3.633 & 0.209 & 2.280/1.091 & 1.600 & 55.578/26.597 & 0.557 & 10.597/5.071\\
H$_2$ & 0.079 & 2.084/0.997 & 0.279 & 3.586/1.716 & 0.142 & 1.387/0.664 & 0.226 & 8.068/3.861 & 0.311 & 5.938/2.841\\
$\rm [CII]$ & 0.065 & 1.500/0.718 & 0.211 & 3.064/1.466 &0.121 & 0.947/0.453 & 0.155 & 5.876/2.812 & 0.235 & 4.839/2.316\\
FUV & 0.060 & 1.151/0.551 & 0.321 & 7.433/3.557 & 0.070 & 1.699/0.813 & 0.118 & 2.955/1.414 & 0.229 & 5.948/2.846\\
K & 0.064 & 1.249/0.598 & 0.199 & 4.699/2.249 & 0.074 & 1.810/0.866 & 0.113 & 2.878/1.377 & 0.180 & 4.818/2.306\\

\hline
\multicolumn{11}{c}{\textbf{Resolution: 2 kpc ($\sigma_{\rm 2D}=851.1\rm\, pc$)}}\\
\hline
Stars & 0.322 & 4.939/2.364 & 0.305 & 6.211/2.972 & 0.331 & 7.184/3.438 & 0.484 & 11.402/5.457 & 0.358 & 8.784/4.204 \\
Gas & 1.277 & 15.934/7.625 & 1.063 &11.807/5.650 & 1.208 & 15.219/7.283 & 2.237 & 28.721/13.744 & 1.725 & 29.105/13.928 \\
H$_2$ & 0.327 & 7.688/3.679 & 0.395 & 5.384/2.576 & 0.339 & 5.680/2.718 & 0.807 & 14.087/6.742 & 0.499 & 10.663/5.103 \\
$\rm [CII]$  & 0.215 & 5.168/2.473 & 0.298 & 4.686/2.243 & 0.255 & 4.385/2.098 & 0.624 & 13.594/6.505 & 0.357 & 8.213/3.930 \\
FUV & 0.252 & 4.381/2.097 & 0.413 & 6.136/2.936 & 0.245 & 5.299/2.536 & 1.008 & 18.717/8.957 & 0.351 & 8.540/4.087 \\
K & 0.250 & 4.208/2.014 & 0.365 & 6.265/2.998 & 0.266 & 5.719/2.737 & 0.926 & 19.595/9.377 & 0.341 & 8.390/4.015 \\

\hline\hline
\end{tabular}
\label{tab:mdyn}
\end{table*}

In general, different gas phases and the stellar component become visible at different wavelengths, traced by different instruments, and we here investigate how the morphology and extent of the evolving galaxy can be assessed using different tracers, i.e. the FIR continuum and [CII] line emissions for gas, and FUV and K-band emissions for stars. The high-redshift Universe is also a very active place: galaxy morphology changes on short timescales, therefore we consider the limitation of observing a limited number of objects at a specific moment in time.

We estimate the FIR emission by integrating a modified Planck spectrum \citep[see, also,][]{decarli18} with the dust temperature obtained from a 3-dimensional table computed by \citet{grassi17} as a function of total gas density $n_{\rm tot} = \rho/(m_{\rm p}\mu)$, temperature, and the visual extinction $A_{\rm V}$, where $\rho$ is the gas mass density, $m_{\rm p}$ is the proton mass, and $\mu$ is the mean molecular weight. 
Instead of performing expensive on-the-fly radiative transfer calculations to estimate $A_{\rm V}$, we follow  \citet{grassi17} and define $A_{\rm V} = [n_{\rm tot}/(10^3\, {\rm cm^{-3}})]^{2/3}$ \citep{glover10,safranekshrader17}. For [CII], we employ the same equation of Paper I \citep{pallottini17a,lupi19b}, where $L_{\rm [CII]}/M_{\rm gas} = \min\{1.0, 0.1[n_{\rm tot}/(10^2\,{\rm cm^{-3})}]\}(Z/{\rm Z_\odot})M_{\rm gas} \rm\, L_\odot\, M_\odot^{-1}$, with $M_{\rm gas}$ the gas mass. For the FUV and K-band emission, we employ the updated \citet{bruzual03} stellar population synthesis models also considered in the simulation, and extract the intrinsic emission in the desired band, without accounting for any absorption along the line-of-sight. In this respect, the reported FUV emission is obviously much higher than the observed one, even if the central MBH was not dominating the emission, whereas the K-band is expected to be much less affected by dust, hence to be more representative of the optical emission we could observe with JWST.

In Fig.~\ref{fig:tracermaps}, we show the quasar host at different redshifts, i.e. from left to right $z=9,8.0,7.8,7.3$, and~7, corresponding to a cosmic time $t_{\rm age}=0.545,0.639,0.660,0.721$, and~0.762~Gyr respectively,
with a field-of-view 20~kpc wide. In the top row, we show the H$_2$ distribution, in the second one the stellar surface density, in the third one the FIR emission, in the fourth one the [CII] flux, and in the bottom ones the unattenuated K-band and FUV emission, respectively. All the maps have been created at a resolution of about 50~pc, to better distinguish the different evolutionary phases. 

From the maps, we clearly observe a very rapid evolution of the gas and stellar distributions in the system, with also two mergers occurring around $z=7.5$. H$_2$ is mostly concentrated in the spiral arms of the disc, especially at $z=9$, when the disc is still strongly perturbed by SNe. At lower redshift, when two infalling satellites approach the galaxy,  H$_2$ can form in moderate amount also along the filaments joining the galaxies. After the merger, the H$_2$ disc appears larger, with spiral arms extending up to a few kpc, although most of the mass is still concentrated within the central kpc.
[CII] closely follows the H$_2$ distribution, with the strongest emission in the spiral arms, although a moderate emission is also present in the low-density gas \citep[see][for a discussion]{lupi20,lupi20b}. Although FIR emission appears more extended, being produced by all the dust within the halo, most of the emission comes from the innermost kpc, where stellar feedback and AGN feedback are able to more effectively heat up the dust \citep[see, also,][]{dimascia21}. However, we have to notice that the only effect of AGN feedback on the dust included in our simple calculations comes from the gas heating, and no radiation has been explicitly included.
In general, for all the gas tracers we observe strong variations in the spatial distribution across `only'  200~Myr, with perturbed discs ($z=9$) alternating to compact structures with tightly wound spiral arms ($z\sim 8$) and then settling again on more extended distributions ($z\lesssim 8$).

The stellar component is less affected by the evolution, with a kpc scale disc always present, surrounded by an evolving low-density environment more susceptible to mergers or any other perturbation. Compared to the intrinsic distribution, the FUV emission is much more compact, consistent with the distribution of star-forming sites. On the other hand, the K-band emission more closely maps the global stellar distribution, hence representing the best tracer of the host galaxy stellar component, even in a young star-forming galaxy like this.

Unfortunately, when degraded at a resolution of 2~kpc, similar to the typical resolution achieved in current observations, all these features and evolutionary phases get smeared out, as can be seen in Fig.~\ref{fig:tracermapslow}, where we report the degraded resolution version of Fig.~\ref{fig:tracermaps}. At this resolution, only the brightest sources are still distinguishable from the background, i.e. the blobs popping out around the central galaxy. In order to check whether these features/companions might be detected, we also report as green dashed contours the $5\sigma$ observation limits for [CII] (ALMA), FUV (NIRCam F115W/F150W2), and K-band (MIRI, F1800W/F21000W) emission, assuming 1 hour integration and 1 square arcsec aperture (for JWST) and 2~kpc resolution (for ALMA). Interestingly, while ALMA would require longer integration for the gas features to be identified, JWST would much more easily detect them, and likely even find smaller features.

In the near future, observations with a few hundred pc resolution will become more common thanks to current and upcoming facilities. As shown in Fig.~\ref{fig:tracermapsmid}, where we report the same maps of Fig.~\ref{fig:tracermaps} degraded at 500~pc, such resolution will allow us to distinguish many of the features that disappeared in the 2~kpc case from the background and the central host. Obviously, in order to observe them, because of their lower brightness, long enough exposure times will be needed \citep[see, e.g.][]{venemans20}. Even more interestingly, because of the spreading over a larger area, some small features get merged together, resulting in an effective higher flux than in the highest resolution map, making them easier to detect \citep[see][for a discussion about this effect in the case of ALMA]{carniani20}.

On a more quantitative level, we report in Table~\ref{tab:mdyn} the dynamical mass of our system as it would inferred by observations, assuming a rotationally-supported disc (first value of each pair) or a dispersion-dominated system (second value of each pair). Compared to \citet{lupi19b}, for the analysis reported here we created 3D data cubes of the field of view, by distributing the luminosity of gas/star particles (according to the observational tracer considered) through a cubic spline kernel along the spatial axes ($x$ and $y$), and a Gaussian profile centred at the particle velocity and with standard deviation the local velocity dispersion along the line of sight. The intrinsic resolution of the cubes has been set to 50~pc and 30~km s$^{-1}$ per channel. To obtain the `observed cubes' we then convolved each velocity channel with a Gaussian kernel of dispersion $\sigma_{\rm 2D}$. 
Finally, we extracted the central spaxel to measure the line profile width, and collapsed the cube into a 2D intensity map to determine the emitting region size. 
In particular, for the line profile, we defined the FWHM of the profile as $1.18(v_{84}-v_{16})$, where $v_{x}$ is the $x$-th percentile of the distribution, whereas for the source size we approximate the `deconvolved size' as $R=\sqrt{R_{\rm fit}^2-\sigma_{2\rm D}^2}$.

As observed in the maps, the quick variations in the structural properties of the system also reflect in the dynamical mass estimates, that instead of  growing monotonically with time (as the real properties), exhibit oscillations of up to a factor of 4-5. 

In conclusion, the extent and morphology of the host galaxy evolve rapidly mostly because substructures get incorporated into the main galaxy. ALMA has already reached 80 mas resolution ($\sim 400$ pc) down to $\sim 50$ $\mu$Jy\,beam$^{-1}$ per 100 km\,s$^{-1}$, leading to a [CII] sensitivity of $1.4\times10^{11}$\,L$_\odot$\,kpc$^{-2}$ in the same velocity bin at $z\sim6.6$ \citep{venemans19}, and could in principle reach $\sim 100$ pc at similar sensitivity. This will allow us to study the warm phase of the gas,
which traces well the main galaxy, but can miss small substructures. The galaxy is most compact in UV where only the central regions of the main galaxy and of the merging substructures are visible. These are the wavelengths at which we can observe today. JWST and MICADO on ELT, which observe at near-IR wavelenghts, will instead have a better ability to trace the full stellar extent to be compared to ALMA’s picture of the gas distribution, and can reach resolutions of $\sim 0.1$ arcsec ($\sim 500$ pc) or better.

\subsubsection{The [CII]--SFR relation}

\begin{figure*}
\centering
\includegraphics[width=\columnwidth]{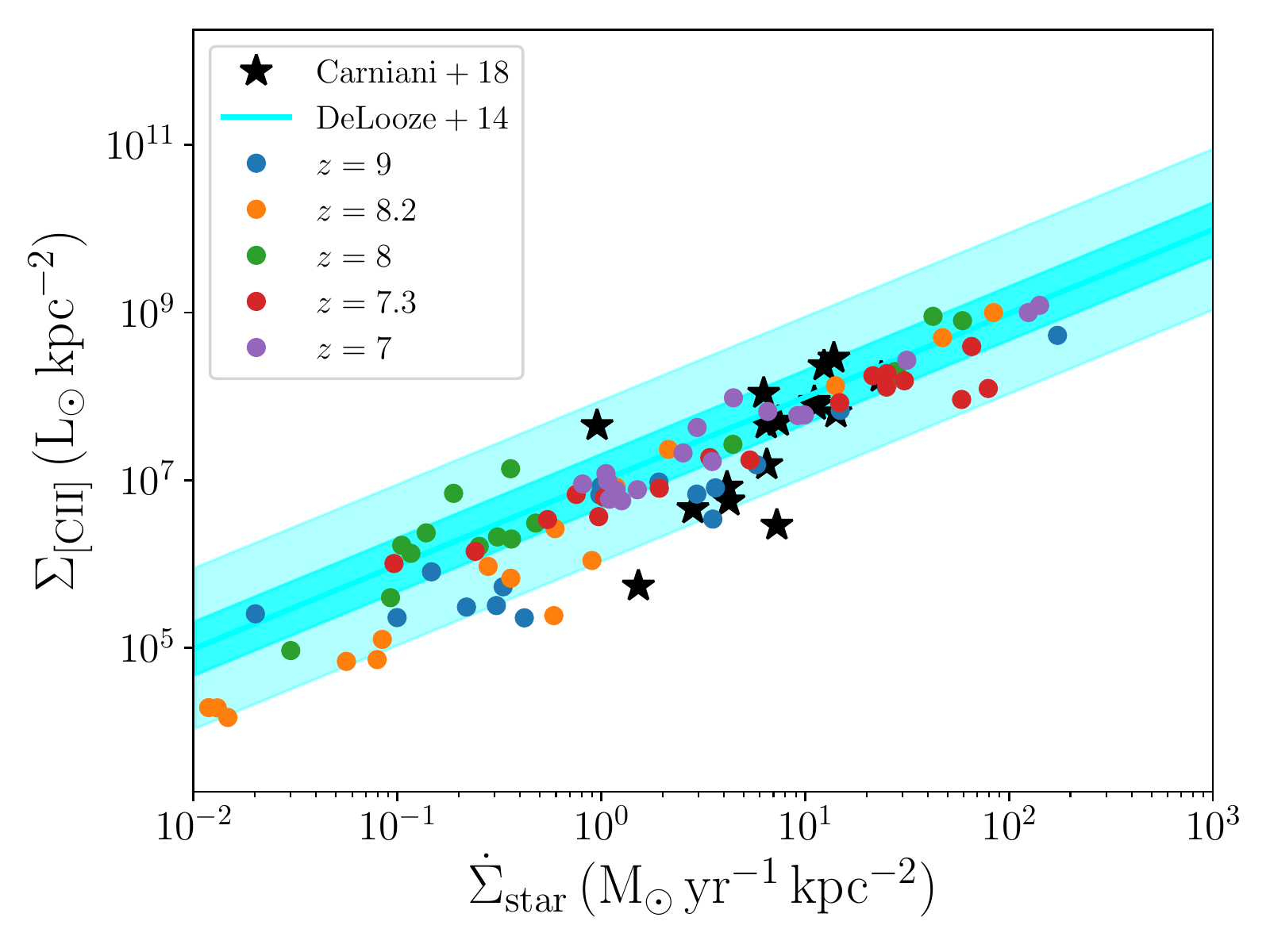}
\includegraphics[width=\columnwidth]{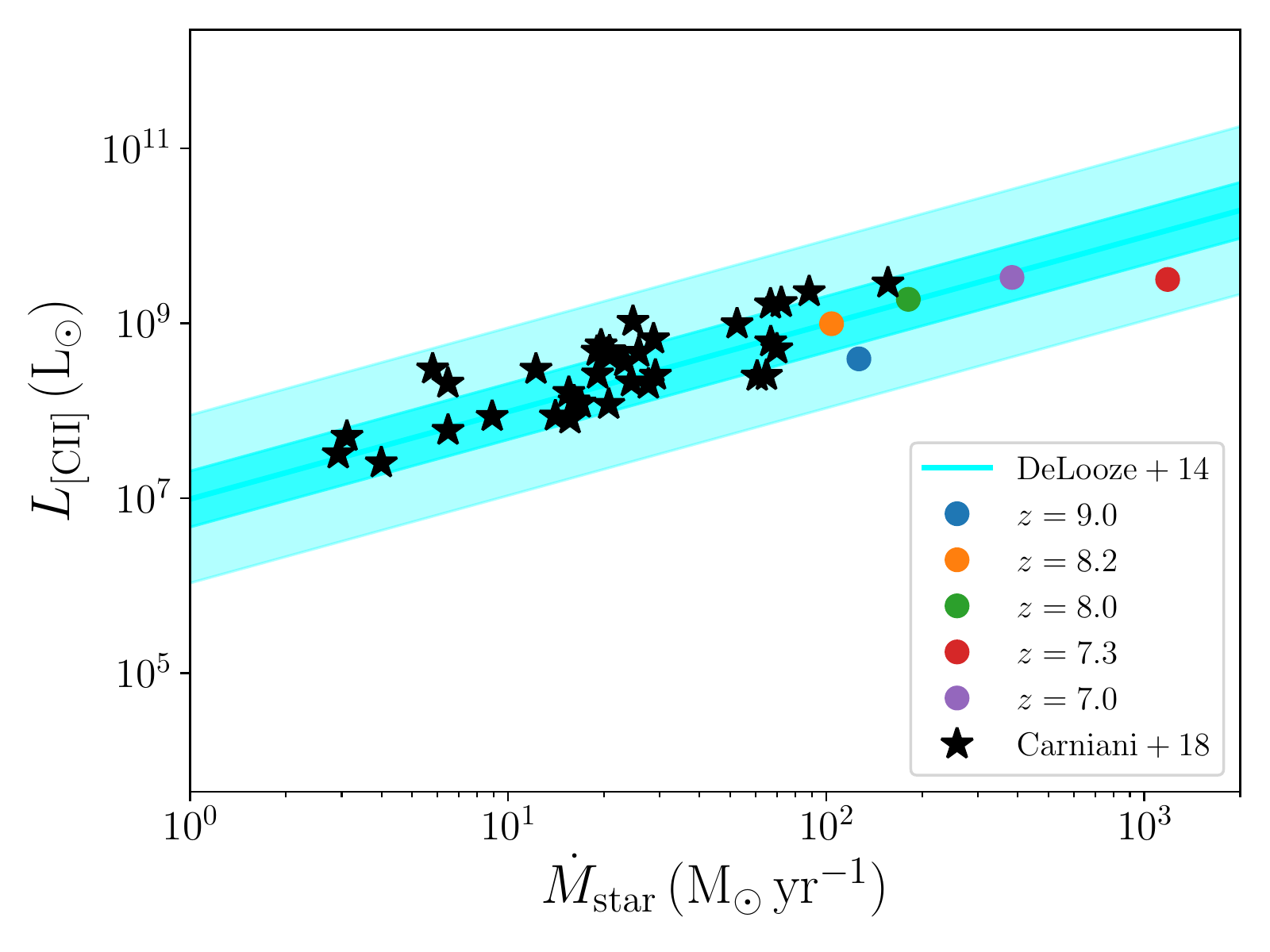}
\caption{[CII]--SFR relation in our simulation (as dots, colour-coded by redshift), compared with the low-redshift sample by \citet[cyan shaded areas, corresponding to $1\sigma$ (darker) and $3\sigma$ scatter (lighter)][]{delooze14} and the high-redshift data by \citet[black stars][]{carniani18}. The left-hand panel corresponds to the resolved relation, estimated by binning the 2D flux maps in 300~pc patches, whereas the right-hand one to the integrated relation. Our results show that, despite small fluctuations of the data relative to the relation at higher redshift, the quasar host well settles on the local relation (either resolved or integrated) and is also in good agreement with the high-redshift galaxies by \citet{carniani18}.}
\label{fig:ciisfr}
\end{figure*}
As we already shown, high-redshift quasar hosts rapidly vary with time their morphology and kinematics, and this can lead to differences in their inferred properties in observations. However, some properties have been shown to be more robust among galaxies, like the correlation between the [CII] luminosity at 158$\mu$m and the SFR, which is valid both at low- and high-redshift \citep{delooze14,herreracamus15,schaerer20}, and independently of the fact that the galaxy is spatially-resolved or not, that suggests [CII] is a good tracer of SF within galaxies, especially in Solar-like conditions \citep{carniani18,schaerer20}. On the other hand, high-redshift observations also showed that many galaxies exhibit a [CII] deficit relative to the FIR luminosity. 
As a consistency check, we compare in Fig.~\ref{fig:ciisfr} the SFR in our quasar host with the [CII] luminosity, as a function of redshift. 

In the left-hand panel, we show the spatially-resolved relation, whereas the integrated one is reported in the right-hand panel. The cyan shaded area corresponds to the local relation by \citet{delooze14}, with the lighter and more extended region corresponding to the 3$\sigma$ uncertainty and the darker and thinner one to $1\sigma$, while the black stars are the high-redshift data by \citet{carniani18}. For the spatially-resolved case, the quasar host exhibits a moderate offset at $8<z\leq 9$ for the low-SF regions \citep[where the metallicity might be moderately lower than solar; see, e.g.][for a discussion]{vallini15,lupi20}, which however disappears at lower $z$, with the galaxy settling perfectly on the local correlation. Our results are also consistent with the high-redshift sources, at all redshifts. As for the integrated correlation, the agreement is preserved, with our sources always lying within the $3-\sigma$ uncertainty of the local relation. Compared to the high-redshift observation by \citet{carniani18}, that targeted normal SF galaxies, the quasar host exhibits a higher SFR, but remain perfectly aligned with the data. This result also confirms the idea that under Solar-like conditions the [CII]--SFR relation is preserved at all redshifts, and our simulation perfectly fits in this picture.

\subsection{The central MBH and its influence radius}
In order to get an accurate estimate of the MBH mass from observations based on gas tracers, we need to resolve the influence region around MBHs, i.e. the region where the gravitational potential of the MBH dominates the gas dynamics. However, since the size of this influence region strongly depends on the (thermo)dynamics of the gas around the MBH, an accurate description of the MBH surroundings on small scales is needed, and is typically unachievable in large scale cosmological simulations. Here, thanks to the detailed chemical network and extremely high mass and spatial resolutions we have, that allow to properly follow the thermodynamic state of the ISM, such an estimate can be performed more accurately. First of all, we determine the influence radius of the MBH in three different ways:
\begin{enumerate}
    \item we compute the Bondi radius $r_{\rm Bondi}={\rm G}M_{\rm BH}/c_s^2$ \citep{bondi52}, which only accounts for the average thermal sound speed of the gas $c_s$ within the MBH smoothing kernel, as if the gas on large scales is at rest, and its energy is fully determined by its internal pressure;
    \item we compute the Bondi-Hoyle-Lyttleton radius, that also includes the average gas-MBH relative velocity $v_{\rm rel}$, which results in less mass actually bound to the MBH, as $r_{\rm BHL}={\rm G}M_{\rm BH}/v_{\rm eff}^2$, where $v_{\rm eff}^2=c_s^2+0.5v_{\rm rel}^2$;\footnote{The factor 0.5 in front of the relative velocity comes from the total energy balance, i.e. 
    \begin{equation}
        \frac{{\rm G}M_{\rm BH}}{r_{\rm BHL}} \sim c_s^2 + \frac{1}{2}v_{\rm rel}^2.
    \end{equation}}
    \item we compare the Keplerian velocity due to the MBH potential with the rotational velocity and the velocity dispersion of the cold gas ($T<10^4$~K) around the MBH, binned in circular annuli, and taking the minimum radius. The hot gas has been excluded to avoid contamination by AGN feedback-affected gas. Moreover, to avoid possible effects of the numerical resolution on the results, we also excluded particles within the MBH force softening length. 
\end{enumerate}

\begin{figure}
\includegraphics[width=\columnwidth,trim=0cm 0cm 0cm 0cm,clip]{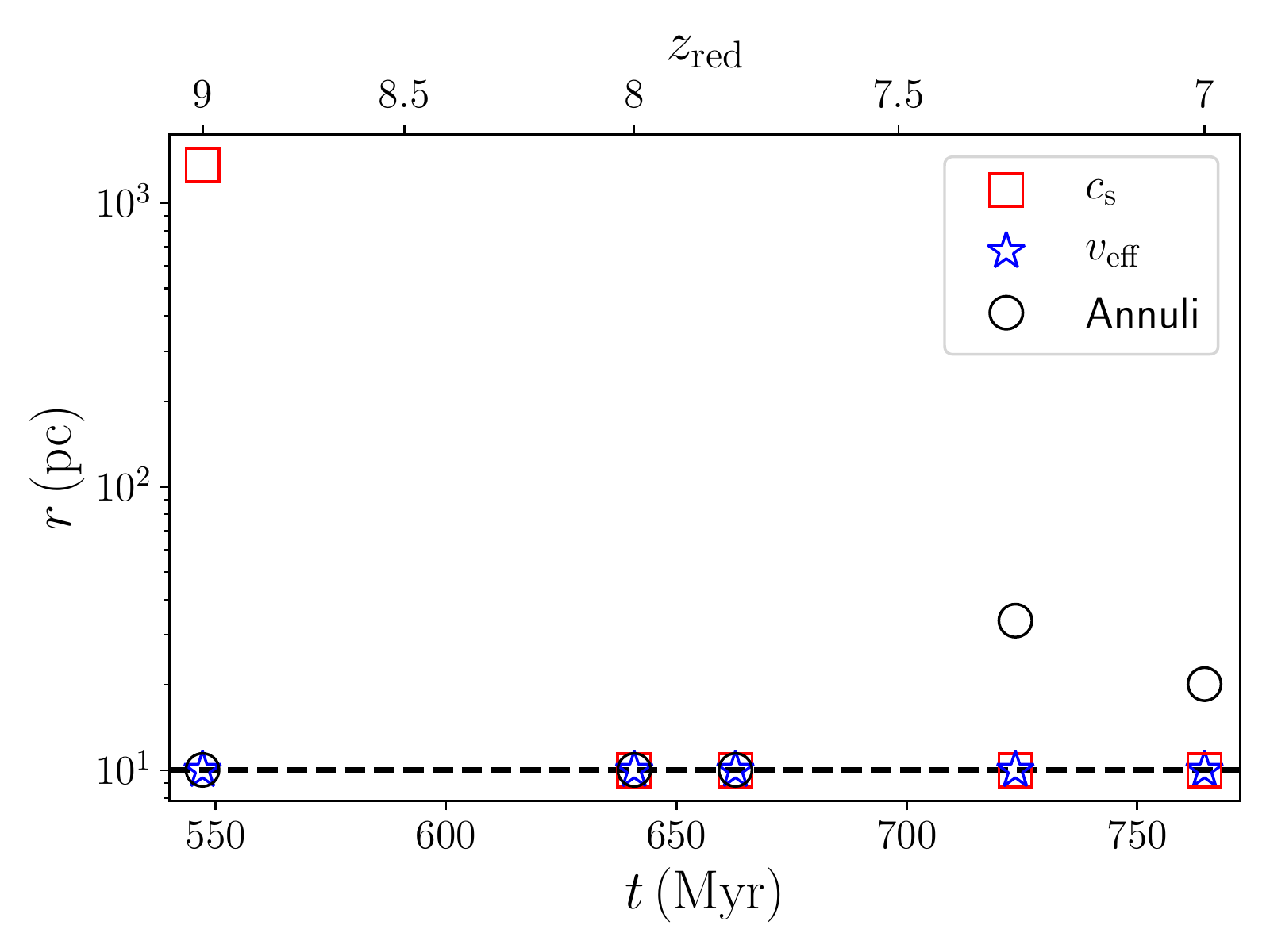}
\caption{Influence radius of the MBH in our simulation at $z=9,8,7.8,7.3$, and 7, estimated as the Bondi radius (red squares), the Bondi-Hoyle-Lyttleton radius (blue stars), and by comparing the Keplerian velocity due to the MBH to the rotational velocity and velocity dispersion of the gas around it (black circles). Apart from some deviations, in particular at high redshift for the Bondi radius, and around $z=7-7.3$ for our third estimator, the radius is always comparable to the spatial resolution of our simulation, hence likely unresolved in current and upcoming observations.}
\label{fig:rinf}
\end{figure}

The results are shown in Fig.~\ref{fig:rinf} at five redshifts. The red squares correspond to $r_{\rm Bondi}$, the blue stars to $r_{\rm BHL}$, and the black circles to our third estimator. The black dashed line corresponds to the MBH softening length. The results show that our estimator results in the BH influence radius being comparable to the BH softening length at high redshift, and then increasing around $z=7$ to about 20--30 pc. On the other hand, the Bondi definition tends to overestimate the influence radius when the MBH feedback is weak (and does not heat up significantly the gas around itself) and the relative gas--MBH velocity is large. Finally, $r_{\rm BHL}$ tends to underestimate the radius when the AGN wind starts to be important, due to its contribution being included in the relative velocity average.

In any case, our results show that, independently of the estimator employed, the influence region around high-redshift MBHs is very small, and this is due to the fact that, although the MBH mass is large, high-redshift systems are quite compact, hence the mass in the central hundred pc is already large and comparable to the MBH mass. This is also consistent with the results by \citet{venemans19}, who found no dynamical signature of the central MBH at resolutions of $\sim 400$~pc, because of the large surface density in the central pixel resulting in a central gas mass about 10 times that of the MBH.

\begin{figure*}
\includegraphics[width=\textwidth]{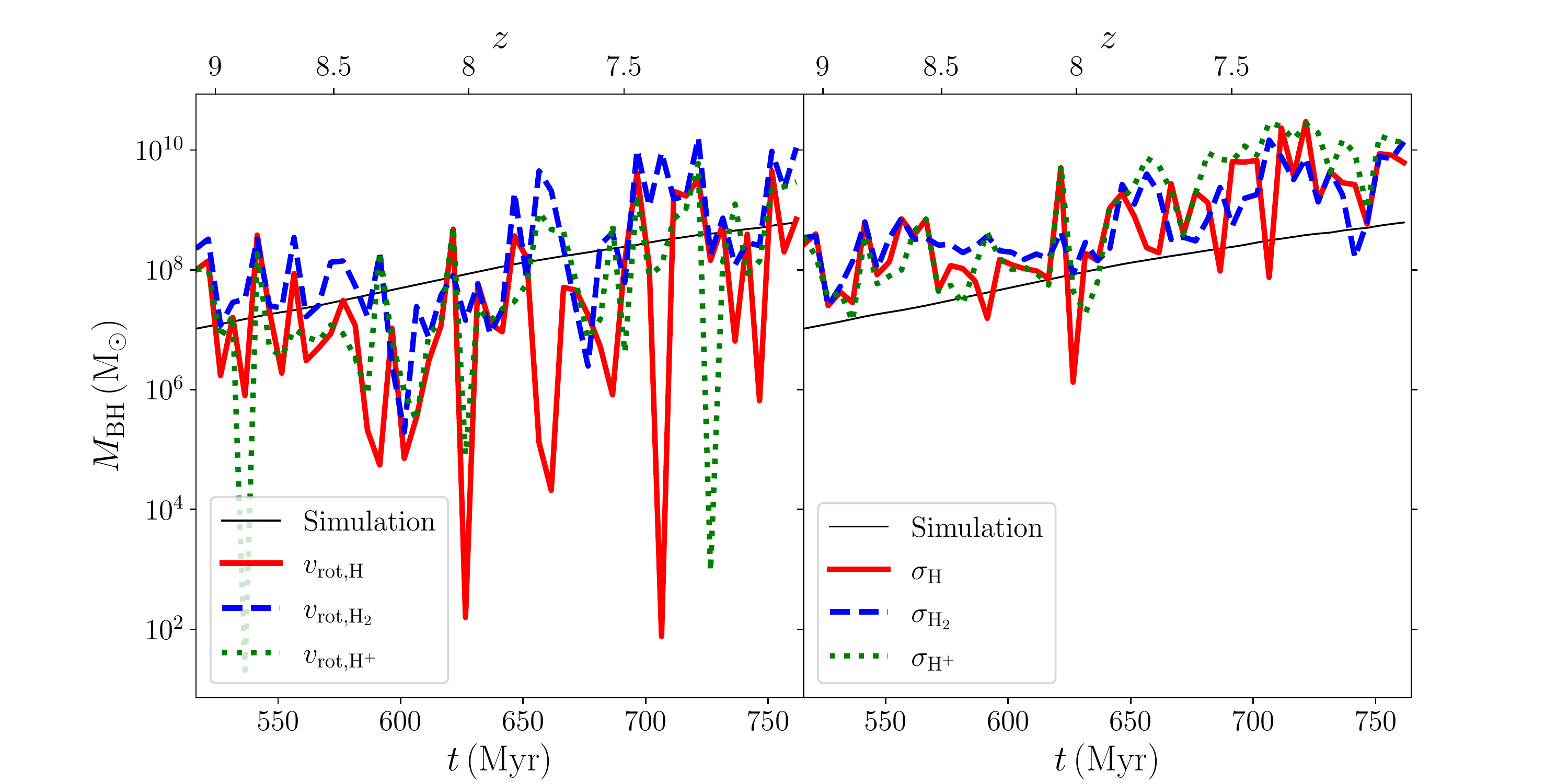}

\caption{Inferred MBH mass assuming the virial theorem, and employing the rotational velocity (left-hand panel) or the velocity dispersion (right-hand panel) of the different gas phases around the MBH (within its kernel, which increases from 10 to 100 pc from $z=9$ to $z=7$). The black thin solid line corresponds to the actual BH mass in the simulation, and the red thick solid, blue thick dashed, and green thick dotted lines to those derived from neutral, molecular, and ionised gas respectively. In the rotationally supported case, the neutral and ionised phases exhibit very strong fluctuation with redshift, and typically underestimate the MBH mass, even by a few orders of magnitude. On the other hand, molecular gas oscillates around the actual mass, with variations up to about one order of magnitude. In the dispersion-dominated case, on the other hand, all tracers tend to overestimate the MBH mass by a factor of a few, up to $\sim 1.3$ order of magnitude.}
\label{fig:bhmass}
\end{figure*}

Nevertheless, it is of utmost importance to assess how accurate the dynamics of different gas phases is in tracing the MBH mass, even if the Keplerian rise is unresolved, to constrain the uncertainty of upcoming observations in the mass estimate. For this reason, in Fig.~\ref{fig:bhmass} we report the MBH estimated from the virial theorem employing the circular velocity (left-hand panel) or the velocity dispersion (right-hand panel) of the different gas phases (molecular, neutral, and ionised) within the MBH smoothing kernel which increases from 10 to 100 pc going from $z=9$ to $z=7$, determined as 
\begin{equation}
    M_{\rm BH,v} = \frac{\langle r\rangle V^2}{\rm G},
\end{equation}
where $\langle r\rangle$ is the density-weighted average distance of the gas particles from the MBH, and $V$ corresponds to the velocity tracer employed. In this case, we removed the relative gas-MBH motion, which is expected to alter the MBH mass estimate increasing the inferred value. The thick lines correspond to the estimate obtained from the neutral (red solid), molecular (blue dashed), and ionised (green dotted) phases, whereas the black thin line is the actual MBH mass in the simulation. We immediately notice that the dispersion-dominated estimate tends to overestimate the inferred MBH mass by a factor of a few, independently of the considered phase. On the other hand, the rotationally-supported case shows extremely large variations for neutral and ionised gas (even by a few orders of magnitude), with a predominant underestimation of the mass, whereas the molecular phase oscillates around the actual value with an error of up to one order of magnitude.

\begin{figure*}
\includegraphics[width=\textwidth]{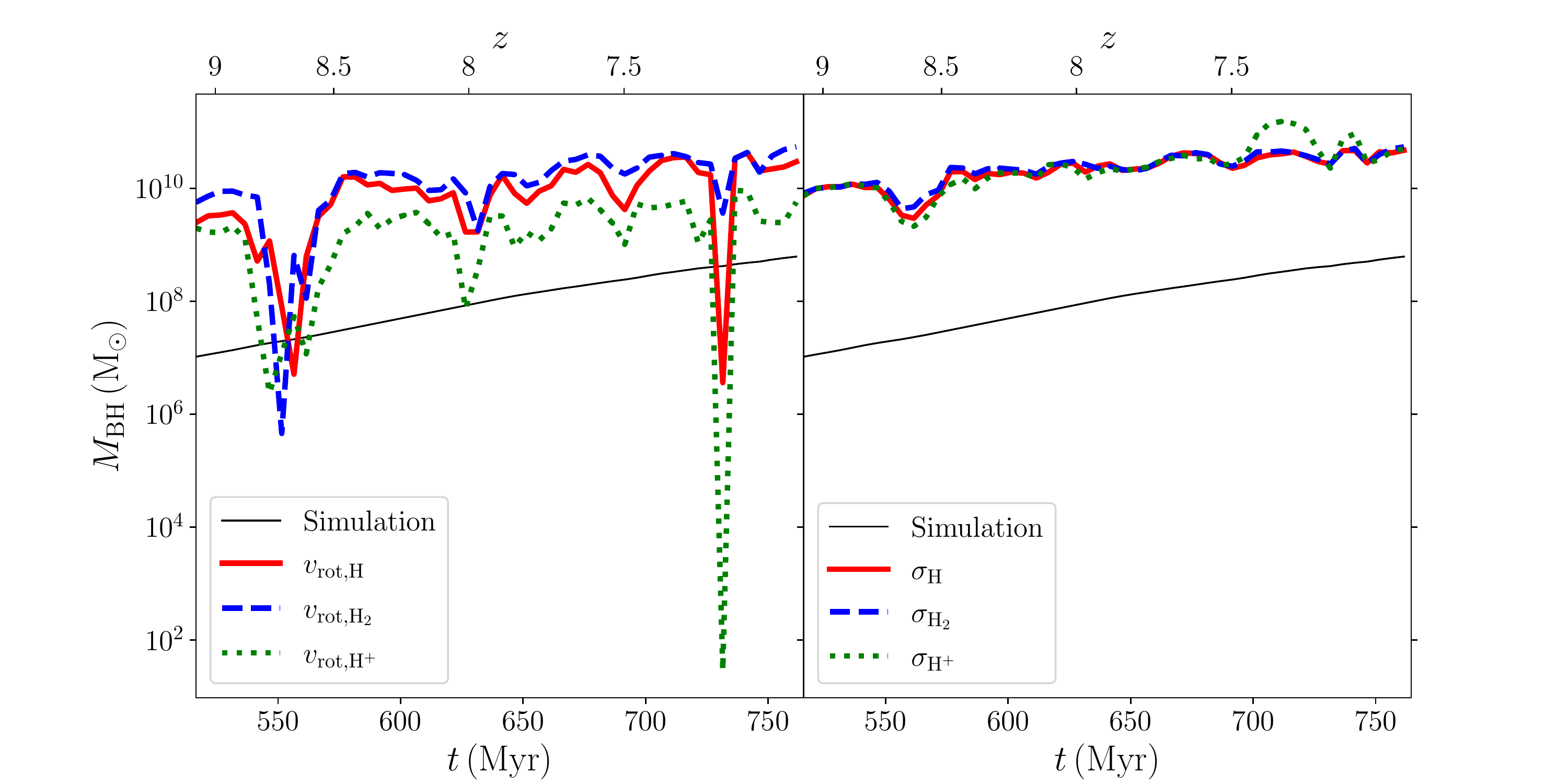}
\caption{Same as Fig.~\ref{fig:bhmass}, but using the velocities within 400~pc from the central MBH. At these scales, the kinematics is already dominated by the central bulge of the galaxy, resulting in a general overestimation of the MBH mass by up to 2 orders of magnitude, independently of the type of support. The only difference can be found in the strong dips in the  rotationally supported case, likely due to momentary phases in which gas is far from a disc-like distribution.}
\label{fig:bhmass400}
\end{figure*}

In Fig.~\ref{fig:bhmass400}, instead, we show the same results when the velocities are averaged over 400~pc, similar to the maximum resolution currently achieved in ALMA observations. In this case, the dynamics is already affected by the stars in the galaxy (notice that the half-mass radius from Fig.~\ref{fig:size_z} is a few hundred pc at $z=7$), hence the inferred mass is up to 2 orders of magnitude larger than the actual MBH mass, independent of the assumption on the kinematics. The only difference when a rotationally-supported structure is assumed is in the two strong dips, that are likely the result of a very perturbed kinematics in the centre of the galaxy.

This result suggests that, even if the influence radius is still beyond the reach of current observations, proper measurements of the molecular gas rotational motion in the central 100~pc around the MBH  allow to estimate with reasonable accuracy (one decade) the MBH mass, assuming a rotationally-supported distribution. If instead only the velocity dispersion can be determined, then the considered phase is almost irrelevant, and the MBH mass can be inferred with reasonable accuracy after a proper calibration of the estimation has been applied (our simulation suggests, for instance, a reduction factor of about 1.2 orders of magnitude on average). However, this is still far from what can be currently achieved ($\sim 400$~pc), and this makes the dynamically inferred MBH masses highly affected by the stellar distribution in the quasar host.

\section{Discussion and conclusions}
\label{sec:conclusions}
In this work, we have addressed the properties of the ISM of high-redshift quasar hosts by means of a high resolution cosmological zoom-in simulation performed with the code \gizmo{} \citep{hopkins15}, described in Paper I. 
Thanks to the on-the-fly non-equilibrium chemistry evolution in the simulation, we have been able to accurately estimate the mass and size evolution of the different phases, ionised, neutral, and molecular with redshift. 

We have also investigated how the quasar host would look like when observed in different bands, both at a fiducial resolution of $\sim 50$~pc and at the typical resolution reached at high-redshift of 2~kpc, showing that important morphological variations occur on very short timescales ($\sim 50-100$~Myr) and some morphological features are smeared out at low resolution, likely changing the system resemblance. 

On a more quantitative level, we also estimated the dynamical mass from various tracers, in order to highlight how well it can be constrained at different redshifts and resolutions. 

An important role in the shaping of the galaxy, and its eventual quenching, is played by the central AGN, that observations suggest is responsible for strong outflows in all phases. In this work, we investigated the relative impact of SNe and the central AGN in producing outflows, and their typical properties, finding that after an initial phase in which SNe dominate (the MBH is not massive enough yet), the AGN takes over and becomes the main driver of the outflows. Nevertheless, the environment in which such systems live is so gas rich and quickly evolving that outflows never dominate over inflows, in any phase. We also found that the molecular outflows far from the galaxy become important only after the AGN takes over, as suggested by \citet{biernacki18}. 

Thanks to our chemical network, we have been able to also assess the reliability of simple temperature cuts compared to a more accurate chemical state selection in the estimate of inflows and outflows properties for the different gas phases, finding that a molecular/neutral transition at $T\sim 300$~K and a neutral/ionised one at $T\sim 3\times 10^4$~K.

Since high-redshift quasar hosts are identified in optical-UV surveys, where the luminosity is dominated by the central source, the presence of high column densities along the line-of-sight could prevent us from properly detecting a large fraction of systems. To assess how many sources we are likely to miss, we evaluated the column density distribution of neutral/molecular hydrogen (exploiting our non-equilibrium chemical network) and the obscuration at different redshifts, finding that such sources are likely to evolve deeply obscured until the galaxy disc settles and the AGN feedback clears out the regions above/below the disc, reducing the obscuration. Nevertheless, a significant fraction ($\sim$ 50-60 per cent) of these sources would still be obscured, hence missed in surveys like SDSS.

Finally, we also assessed whether accurate MBH mass measurements could be achieved in the near future in these systems, by means of high-resolution observations approaching scales comparable to the influence region around the MBH. Despite the influence region being still beyond reach for the capabilities of current and upcoming facilities for the near future, independently of the estimator used, a suitable calibration of mass estimators based on the virial theorem and different gas phases (molecular gas is in general the best phase that can be used) is able to yield reasonable estimates, within about one order of magnitude uncertainties.

\section*{Acknowledgements}
We thank the anonymous referee for their useful comments that improved the quality of the manuscript. AL acknowledges funding from MIUR under the grant PRIN 2017-MB8AEZ.
SB acknowledges financial support from Millenium Nucleus NCN19\_058 (TITANs).
This work was granted access to the High Performance Computing resources of CINES under the allocations A0020406955, and A0040406955 by GENCI, and it has made use of the Horizon Cluster, hosted by Institut d'Astrophysique de Paris, for the analysis of the simulation results.
The maps reported in this work have been created using \textsc{pynbody} \citep{pynbody}.

\section*{Data Availability Statement}
The data underlying this article will be shared on reasonable request to the corresponding author.

\bibliographystyle{mnras}
\bibliography{./Biblio}

\label{lastpage}
\end{document}